\documentclass[aps,pre,superscriptaddress,reprint,amsfonts,amsmath]{revtex4-1}

\usepackage[final]{graphicx}

\usepackage{mathtools}

\usepackage{comment}
\usepackage[utf8]{inputenc}

\usepackage[usenames,dvipsnames]{xcolor}
\usepackage[normalem]{ulem}
\newcommand\TODO[1]{%
  {\fboxsep=.3ex\colorbox{BrickRed}{\textcolor{white}{\textsc{TODO}}}
  \textcolor{BrickRed}{#1}}%
}

\def\TODO#1{\unskip}

\DeclareMathOperator{\smoluch}{\Omega}

\newcommand{\tlname}[1]{\ensuremath{\mathit{#1}}}
\DeclareMathOperator\projP{\mathcal P}
\DeclareMathOperator\projQ{\mathcal Q}
\let\bs\boldsymbol
\def\delrho{\tlname{\delta\varrho}}

\DeclarePairedDelimiter{\lrb}{(}{)}

\DeclareMathOperator{\smol}{\Omega}
\DeclareMathOperator{\dsmol}{\delta\Omega}
\DeclareMathOperator{\adsmol}{\smol^{\dag}}
\DeclarePairedDelimiter{\avrg}{\langle}{\rangle}
\DeclarePairedDelimiter{\proj}{\rangle}{\langle}
\DeclareMathOperator{\adsmolt}{\smol_t^{\dag}}
\DeclareMathOperator{\adsmoltirr}{\smol_{t,\text{irr.}}^{\dag}}

\begin{document}
\title{Mode-Coupling Theory for Active Brownian Particles}
\date{\today}
\def\dlr{\affiliation{Institut f\"ur Materialphysik im Weltraum,
  Deutsches Zentrum f\"ur Luft- und Raumfahrt (DLR), 51170 K\"oln,
  Germany}}
\def\hhu{\affiliation{Department of Physics,
  Heinrich-Heine Universit\"at D\"usseldorf,
  Universit\"atsstr.~1, 40225 D\"usseldorf, Germany}}

\author{Alexander Liluashvili}\dlr
\author{Jonathan Onody}\dlr
\author{Thomas Voigtmann}\dlr\hhu

\begin{abstract}
We present a mode-coupling theory (MCT) for the high-density dynamics of
two-dimensional spherical active Brownian particles (ABP).
The theory is based on the integration-through-transients (ITT) formalism
and hence provides a starting point for the calculation of non-equilibrium
averages in active-Brownian particle systems.
The ABP are characterized by a self-propulsion velocity $v_0$, and by their
translational and rotational diffusion coefficients, $D_t$ and $D_r$.
The theory treats both the translational and the orientational degrees of
freedom of ABP explicitly. This allows to study the effect of
self-propulsion of both weak and strong persistence of the swimming
direction, also at high densities where the persistence length
$\ell_p=v_0/D_r$ is large compared to the typical interaction length scale.
While the low-density dynamics of ABP is characterized by a single
P\'eclet number, $\tlname{Pe}=v_0^2/D_rD_t$, close to the glass transition
the dynamics is found to depend on $\tlname{Pe}$ and $\ell_p$ separately.
At fixed density, increasing the self-propulsion velocity causes structural
relaxation to sped up, while decreasing the persistence length slows down
the relaxation.
The theory predicts a non-trivial idealized-glass-transition diagram in the three-dimensional
parameter space of density, self-propulsion velocity and rotational diffusivity.
The active-MCT glass is a nonergodic state where correlations of initial
density fluctuations never fully decay, but also an infinite memory of
initial orientational fluctuations is retained in the positions.
\end{abstract}

\maketitle

\section{Introduction}

The collective dynamics of self-propelled particles and the
related behavior of dense suspensions of microswimmers
have received an increasing amount of attention in the past
few years.
The physical principles of this dynamics are relevant for many biophysical
questions. For example, mechanisms at work in wound healing
(tissue repair) have been likened to the collective dynamics found in
model microswimmer systems \cite{biophys:Poujade.2007,biophys:Petitjean.2010,biophys:Trepat.2009,biophys:Angelini.2010,biophys:Bi.2016} \TODO{and papers I used in Greifswald talk}.
The collective dynamics of bacteria colonies \TODO{check}
\cite{biophys:Peruani.2012},
as well as the dynamics of the cell cytoskeleton show slow dynamics typically
associated with that arising from crowding effects at high densities and
associated glass-like slowing down \cite{biophys:Fabry.2001,biophys:Bursac.2005,biophys:Wang.2011b}.

Colloidal self-propelled particles provide paradigmatic model systems to study
the qualitative effects of swimming, and more generically the generic
features of a broad class of intrinsically non-equilibrium matter.
In experiment, self-phoretic Brownian particles provide a good realization
of this model.
Half-capped ``Janus particles'' provide a surface-mediated mechanism to
convert energy provided by light fields or chemical fuel into directed
motion, superimposed on the Brownian translational and rotational diffusion
of passive colloids
\cite{rheology:Erbe.2008,biophys:Baraban.2012,biophys:Palacci.2010,biophys:Buttinoni.2013,biophys:Volpe.2011}.\TODO{cite Theurkauff.2012?}

One of the simplest theoretical models in this context
is that of active Brownian particles (ABP)
\cite{biophys:Zoettl.2016,biophys:Marchetti.2016}. These are
orientable colloidal particles that undergo ``passive'' translational and
rotational Brownian motion (ignoring
hydrodynamic interactions in the following), and in addition an ``active'' drift
along a body-fixed orientation axis due to intrinsic self-propulsion
forces. In two spatial dimensions, particles are described by their
positions $\vec r_i$ and orientation angles $\theta_i$ with respect to
a fixed direction in space. Here, $i=1,\ldots N$ labels the particles.
The equations of motion read, in the case of spherically symmetric
interaction forces,
\begin{subequations}\label{eq:abpsde}
\begin{align}
d\vec r_i&=\vec F_i/\zeta\,dt+\sqrt{2D_t}\,d\vec W_i+v_0\vec o(\theta_i)\,dt\,,\\
d\theta_i&=\sqrt{2D_r}\,dW_{\theta_i}\,,
\end{align}
\end{subequations}
where $\vec o(\theta)=(\cos\theta,\sin\theta)^T$ is the orientation vector.
We will abbreviate $\vec o_i=\vec o(\theta_i)$. The elements of $dW$ are
independent Wiener processes, and the $\vec F_i$ are the interaction forces.
The friction coefficient $\zeta=1/\beta D_t$ is taken to obey the
fluctuation-dissipation theorem for the translational Brownian motion of
the passive-particle system, with inverse temperature $\beta$.
The key parameters characterizing the dynamics of a single
spherical ABP are $D_t$, $D_r$, and $v_0$: the translational short-time
diffusion coefficient, the rotational diffusion coefficient, and the
self-propulsion velocity.
Although for a passive colloid, $D_r$ and $D_t$ are coupled due to
the hydrodynamics of the solvent, for spherical ABP it makes sense to treat
$D_r$ as an independent model parameter. Different self-propulsion mechanisms
may impose different persistence effects on the orientation, and to some
extent these can be captured by varying $D_r$ \cite{biophys:Ghosh.2015}.

We focus on spherical ABP with strongly repulsive interactions, modeled
by the equilibrium structure of a hard-sphere suspension. At high densities,
such systems (with suitable size polydispersity) are known from simulation
to form
glasses \cite{biophys:Ni.2013,biophys:Fily.2013,biophys:Kuan.2015},
as do related active-particle models
\cite{biophys:Berthier.2013,biophys:Berthier.2014,biophys:Levis.2015,biophys:Szamel.2015,biophys:Mandal.2016,biophys:Marchetti.2016,biophys:Bi.2016}.

The theory of ABP has been extensively studied at low and moderate densities
\TODO{cite Franosch, seek some others} \cite{biophys:Takatori.2014,biophys:Takatori.2016,biophys:Yan.2015,biophys:Yan.2015bpre,biophys:Kurzthaler.2016}.
At large time scales, $D_rt\gg1$, an ``effective-diffusion limit'' can be
performed to map the dynamics of the dilute system onto
Brownian motion with an effective diffusion coefficient:
$D_\text{eff}=D_t+D_\text{act}=D_t(1+\tlname{Pe}/2)$, where
$\tlname{Pe}=v_0^2/D_rD_t$ defines the relevant P\'eclet number
and $D_\text{act}=v_0^2/2D_r$ is an activity-induced diffusivity.
Accounting for this enhancement of diffusivity, many properties of dilute
ABP suspensions and their phase behavior can be explained
\cite{biophys:Takatori.2014,biophys:Takatori.2016}.
The mapping however requires that all relevant length scales in the
problem are large compared to the average swim length $\ell_p=v_0/D_r$,
and that one probes the system on length scales larger than that
\cite{biophys:Yan.2015,biophys:Yan.2015bpre}.

The high-density dynamics of ABP is less well explored.
To describe glassy behavior of active systems, theories of the
glass transition have been extended from the passive near-equilibrium
case to include self propulsion \cite{biophys:Wang.2011,biophys:Farage.2014pre,biophys:Farage.2015,biophys:Szamel.2015,biophys:Szamel.2016,biophys:Nandi.2016pre,biophys:Nandi.2016bpre}, using various different models and approximations.
One particular reference point for the passive Brownian system
is the mode-coupling theory of the glass transition (MCT), both in
3D \cite{glass-theory:Goetze.2009} and in 2D \cite{glass-theory:Bayer.2007}.
MCT has been extended to deal with spherical ABP in the effective-diffusion
limit \cite{biophys:Farage.2014pre}. In this limit,
the equations of motion, Eq.~\eqref{eq:abpsde}, can be formally
reduced to eliminate the orientation angles $\theta_i$ as explicit variables,
in a procedure akin to the well-known reduction of the phase-space
Langevin equation to the configuration-space Brownan dynamics.
One obtains $d\vec r_i=\vec F_i/\zeta_\text{eff}+\sqrt{2D_t}\,d\vec W_i$
where $\zeta_\text{eff}$ \TODO{$\zeta_\text{eff}=xxx$} is an effective friction coefficient.
The resulting MCT takes the same form as the passive-equilibrium theory,
with an activity-dependent prefactor in the memory kernel that reflects
the non-equilibrium nature of the system through a violation of the
fluctuation-dissipation theorem, $\zeta_\text{eff}\neq1/\beta D_t$.
Consequently, activity enters this theory only through the P\'eclet
number $\tlname{Pe}$.
This MCT approach extends easily to mixtures of active and passive
particles \cite{biophys:Ding.2015pre}. It predicts a shift of the glass
transition to higher densities, in qualitative agreement with simulation
results.

However, the approach to a glass transition implies transient caging
of particles on a length scale $\ell_c\sim0.1\sigma$ (where $\sigma$ is
a typical particle size), over increasingly long times. It is not evident
that the effective-diffusion approach remains valid as the condition
$\ell_c<\ell_p$ is easily violated for typical swim speeds used in
simulation and experiment. This calls for a theoretical treatment that
starts directly from Eq.~\eqref{eq:abpsde}, rather than from a further
reduced description of the dynamics.

\TODO{A MCT-based theory similar in spirit was recently derived for the
AOUP check where to sort this in -- cite Soft Matter paper.
There, a near-equilibrium assumption was implicitly made.}

Besides the ABP, the best studied model
of active particles to date is the one proposed by Berthier, Szamel, and
coworkers \cite{biophys:Szamel.2015,biophys:Szamel.2016},
that of active Ornstein-Uhlenbeck particles (AOUP). Here, particles are
described by their positions and an activity vector that represents
the swimming direction and evolves according to an Ornstein-Uhlenbeck
process \TODO{Check how} providing colored noise for the evolution of the
positions.
The model thus belongs to a class where activity is modeled as leading to
a persistent random walk of the particles \cite{biophys:Szamel.2014,biophys:Marconi.2015,biophys:Sadjadi.2015}.
This and related models have been studied in simulation \cite{biophys:Berthier.2014,biophys:Levis.2014,biophys:Levis.2015,biophys:Mandal.2016,biophys:Flenner.2016pre} as the extension
of paradigmatic passive glass formers to the non-equilibrium regime.
A theoretical treatment based on MCT was established \cite{biophys:Szamel.2016},
under the simplifying assumption that the particle positions evolve on a
time scale larger than the time scale governing the evolution of the
activity vector. This is not unlike the effective-diffusion
approximation made in
previous studies of high-density ABP systems. Predictions of the
AOUP-MCT have been tested in computer simulation \cite{biophys:Flenner.2016pre}.
In the ``athermal'' case where activity is the sole driving force,
the glass transition was found to depend non-monotonically on the
activity strength \cite{biophys:Szamel.2016}. This appears to be different
for ``thermal'' ABP, according to present simulation results.
Thus the connection among different models of active colloidal systems
remains to be studied in more detail.

In this paper, we develop a MCT of spherical ABP that
treats both the positional and orientational degrees of freedom on
equal footing.
This avoids the reduction to a near-equilibrium or an effective-diffusion
description, and allows to study the qualitative effects of
self-propulsion of various persistence lengths in the high-density regime.
In particular it allows to study the limits $D_r\to0$ and $D_r\to\infty$
as interesting reference cases \cite{biophys:Reichhardt.2014} that
provide valuable insight into the mechanisms by which swimming
modifies the caging dynamics; 
Our approach is based on the integration-through transients (ITT)
formalism as a formal approach to deal with the self-propulsion force
as an arbitrarily strong perturbation of the passive-equilibrium dynamics.

The paper is organized as follows: in Sec.~\ref{sec:mct} we derive the
MCT for two-dimensional spherical ABP, including rotational degrees of
freedom. Section~\ref{sec:dynamics} shows numerical results for the
dynamical density correlation functions close to the glass transition;
Sec.~\ref{sec:glasstransition} is devoted to a discussion of the
dependence of the glass-transition point on activity.
Section~\ref{sec:conclusions} concludes.

\section{Mode-Coupling Theory}\label{sec:mct}

The statistical information of the dynamics of a system composed of $N$
spherical ABP is encoded
in the Smoluchowski equation for the configuration-space distribution
function $p(\Gamma,t)$, i.e., the probability density that corresponds to
the Markov process described by the stochastic differential equations
Eq.~\eqref{eq:abpsde}. Here $\Gamma=(\Gamma_r,\Gamma_\theta)
=(\vec r_1,\ldots,\vec r_N,\theta_1,\ldots,\theta_N)$ labels points in
configuration space. There holds $\partial_tp(\Gamma,t)=\smol(\Gamma)
p(\Gamma,t)$ with the Smoluchowski operator
\begin{multline}\label{eq:smoluchowski}
  \smol = \sum_{j = 1}^N {D_t \vec{\nabla}_j \cdot \lrb*{\vec{\nabla}_j - \beta \vec{F}_j} + D_r \partial_{\theta_j}^2 - v_0 \vec{\nabla}_j \cdot \vec{o}(\theta_j) }\\
	 = \smol_\text{eq}(D_t, D_r) + \dsmol(v_0)\,,
\end{multline}
where $\dsmol(v_0)=-v_0\sum_j\vec\nabla_j\cdot\vec o(\theta_j)$ is
the term that represents the active motion.
The interaction forces are assumed to follow from a spherically symmetric
interaction potential,
$\vec F_j=-\vec\nabla_jU(\Gamma_r)$.

Equation~\eqref{eq:smoluchowski} for $v_0=0$ describes the passive Brownian
system. It admits the equilibrium solution $p_\text{eq}(\Gamma_r)\propto
\exp[-\beta U(\Gamma_r)]$.
The integration-through transients (ITT) formalism expresses averages of
observables in the non-equilibrium system through history integrals involving
transient correlation functions, i.e., correlation functions that contain
the full non-equilibrium time evolution but are taken with the equilibrium
distribution function. The starting point of ITT is the identity
$\exp[\smol t]=1+\int_0^tdt'\exp[\smol t']\smol$. Using this
identity to rewrite $p(t)=\exp[\smol t]p(0)$, and assuming that
for $t=0$ the system starts in equilibrium, one gets a generalized
Green-Kubo formula for any observable $A$. For the special case of the
spherical ABP system,
\begin{equation}\label{eq:itt}
  \langle A\rangle_t=\langle A\rangle_\text{eq}
  -\beta v_0\int_0^tdt'\left\langle\sum\nolimits_{j=1}^N\vec o_j\cdot\vec F_j\,
  e^{\smol^\dagger t'}A\right\rangle_\text{eq}\,.
\end{equation}
The desire to evaluate expressions like this prompts the development of
a theory of transient correlation functions.
In the following, we will drop the subscript ``eq'' and implicitly
perform all averages over the equilibrium distribution.

The collective motion of the particles is described by the local
density fluctuations, $\varrho(\vec r,\theta)=\sum_{j=1}^N
\delta(\vec r-\vec r_j)\delta(\theta-\theta_j)$, respectively their
Fourier transform
\begin{equation}
  \delrho_l(\vec q)=\sum_{j=1}^Ne^{i\vec q\cdot\vec r_j}e^{il\theta_j}
  /\sqrt N
  \,.
\end{equation}
with integer angular indices $l=-\infty,\ldots,\infty$.
We assume the system to remain in a homogeneous, translationally invariant
and isotropic state. Then, the equilibrium static structure factor matrix
depends on the wave vector only through $q=|\vec q|$,
\begin{equation}
  S_{ll'}(q)=\langle\delrho_l^*(\vec q)\delrho_{l'}(\vec q)\rangle\,.
\end{equation}
Also the two-point
density correlation functions are diagonal in wave-vector space under these
conditions. Since the interaction potential is spherically symmetric,
the matrix $\bs S(q)$ takes the simple form
\begin{equation}
  S_{ll'}(q)=\delta_{ll'}(1+\delta_{l0}(S_q-1))\,,
\end{equation}
i.e., it is the unit matrix with its $(00)$ element replaced by $S_q$,
the usual static structure factor of the equilibrium system of spherical
particles.

The time-dependent transient density correlation functions are defined as
\begin{equation}
  S_{ll'}(\vec q,t)=\left\langle\delrho_l^*(\vec q)
  e^{\smol^\dagger t}\delrho_{l'}(\vec q)\right\rangle\,.
\end{equation}
Here, the adjoint (or backward) Smoluchowski operator provides the
temporal evolution,
\begin{equation}
  \smol^\dagger=
	\sum_{j = 1}^N D_t \lrb*{\vec{\nabla}_j + \beta \vec{F}_j} \cdot \vec{\nabla}_j + D_r \partial_{\theta_j}^2 + v_0 \vec{o}_j \cdot \vec{\nabla}_j \,.
\end{equation}
We use the convention that this operator acts to everything on its right,
but not on the distribution function itself.
There holds $\bs S(\vec q,0)=\bs S(q)$.
\TODO{check:} The normalized
correlator is defined by $\bs\Phi(\vec q,t)=
\bs S(\vec q,t)\cdot\bs S^{-1}(q)$.

The correlation functions are defined in a specific laboratory frame of
reference, with respect to which particle orientations are measured.
For this reason, the correlation functions depend a~priori on the
direction of the wave vector $\vec q$. However, there hold simple
transformation rules to transform the correlators to a rotated reference
frame. Consider a rotation around an angle $\alpha$,
$\vec r\mapsto\vec r^{\,\prime}=\bs D(\alpha)\cdot\vec r$ and $\theta\mapsto
\theta'=\theta+\alpha$, witth $\bs D\cdot\bs D^T=1$ a rotation matrix.
This changes
$\delrho_l(\vec q)\mapsto\delrho_l(\vec q^{\,\prime})\exp[il\alpha]$
where $\vec q^{\,\prime}=\bs D\cdot\vec q$.
The transformation is thus given by a unitary representation $\bs u(\alpha)$
of the orientation group $\tlname{SO}(1)$, given by
$u_{ll'}=\exp[-il\alpha]\delta_{ll'}$. (There holds $\bs u(\alpha)
\cdot\bs u(\beta)=\bs u(\alpha+\beta)$, $\bs u(\alpha)\bs u^\dagger(\alpha)
=\bs1$, $\bs u(0)=\bs1$, and $\bs u(\alpha+2\pi)=\bs u(\alpha)$, as well
as $\det\bs u=1$.)
One easily shows that the Smoluchowski operator itself is invariant under
rotation, $\smoluch(\Gamma)=\smoluch'(\Gamma')$, seperately in all its terms.
(To see this, recall $\vec\partial^{\,\prime}\cdot\vec o(\theta')
=(\bs D^{-1}\cdot\vec\partial)\cdot\vec o(\theta+\alpha)
=\vec\partial\cdot\bs D\cdot\vec o(\theta+\alpha)
=\vec\partial\cdot\vec o(\theta)$.
Under rotation, the equilibrium distribution function remains invariant,
and thus one obtains the transformation rule
\begin{equation}\label{eq:rotation}
  \bs S(\vec q,t)\mapsto\bs u(\alpha)\cdot\bs S(\vec q^{\,\prime},t)
  \cdot\bs u^\dagger(\alpha)\,.
\end{equation}
We will make use of this relation to restrict the discussion of the
correlation functions to wave vectors aligned with a particular
spatial direction, chosen by $\vec q=q\vec e_y$. Note that
Eq.~\eqref{eq:rotation} confirms that $S_{00}(q,t)$ is in fact invariant
under rotations.

An equation of motion for $\bs S(\vec q,t)$ can be derived using the
Mori-Zwanzig projection operator formalism.
Introduce the projection operator onto density fluctuations,
\begin{equation}
  \projP=\sum_{l_1l_2}\delrho_{l_1}(\vec q)\rangle
  S^{-1}_{l_1l_2}(q)\langle\delrho^*_{l_2}(\vec q)\,,
\end{equation}
and set $\projQ=1-\projP$.
One now writes $\partial_t\exp[\adsmol t]=\adsmol(\projP+\projQ)\exp[\adsmol t]$
and rewrites the second term using the Dyson decomposition
\begin{equation}
  e^{\adsmol t}=e^{\adsmol\projQ t}+\int_0^tdt'\,e^{\adsmol\projQ(t-t')}
  \adsmol\projP e^{\adsmol t'}
\end{equation}
to obtain
\begin{multline}\label{eq:mz1}
  \partial_t\bs S(\vec q,t)=-\bs\omega(\vec q)\cdot\bs S^{-1}(q)
  \cdot\bs S(\vec q,t)
  \\ +\int_0^tdt'\,\bs K(\vec q,t-t')\cdot\bs S^{-1}(q)
  \cdot\bs S(\vec q,t')
\end{multline}
where $\bs\omega(\vec q)$ generalizes the collective diffusion matrix,
\begin{multline}
  \omega_{ll'}(\vec q)=-\left\langle\delrho^*_l(\vec q)\smol^\dagger
  \delrho_{l'}(\vec q)\right\rangle
  \\
  =\lrb*{q^2 D_t + l^2D_r}\delta_{ll'}
  -\frac{iqv_0}2e^{-i(l-l')\vartheta_q}S_{ll}(q)\delta_{|l-l'|,1}\,,
\end{multline}
writing
$\vec q=q(\cos\vartheta_q,\sin\vartheta_q)^T$.
The memory kernel $\boldsymbol K(\vec q,t)$ is given by
\begin{equation}\label{eq:memk}
  K_{ll'}(\vec q,t)=\left\langle\delrho_l^*(\vec q)\adsmol\projQ
  e^{\projQ\adsmol\projQ t}\projQ\adsmol\delrho_{l'}(\vec q)\right\rangle
\end{equation}
It describes the renormalization of the diffusion matrix due to many-body
interactions.

\TODO{At low densities, the memory kernel can be dropped, can we show this
generically? Then, $\partial_t\boldsymbol\Phi=-\boldsymbol\omega\cdot
\boldsymbol\Phi$ can be solved formally. This reproduces the solution
of Kurzthaler and Franosch.}
For small density, $\rho=N/V\to0$ (where $V$ is the system volume),
one can drop the memory kernel in Eq.~\eqref{eq:mz1}. The formal
solution, $\bs S(\vec q,t)=\exp[-\bs\omega(\vec q)t]$, reproduces the
exact solution \TODO{in terms of spheroidal wave functions and their
2D analog the Mathieu functions} as discussed in 3D \cite{biophys:Kurzthaler.2016}.

The slowing down of the dynamics close to a glass transition is driven
by slow positional density fluctuations. This suggests to split
the time-evolution operator, $\smol=\smol_T(D_t,v_0)+\smol_R(D_r)$.
The matrix elements of the translational and rotational parts will be
written as $\bs\omega(\vec q)=\bs\omega_T(\vec q)+\bs\omega_R(\vec q)$,
i.e., $\omega_{R,ll'}(\vec q)=l^2D_r\delta_{ll'}$.
Correspondingly, we decompose the memory kernel into four contributions,
$\bs K(t)=\bs K^{TT}(t)+\bs K^{TR}(t)+\bs K^{RT}(t)+\bs K^{RR}(t)$,
given by replacing the operators $\adsmol$ appearing to the right and to the
left of the reduced propagator in Eq.~\eqref{eq:memk} by their decompositions.

Since in our model of spherical ABP, the rotational degrees of freedom
never slow down, all contributions to the memory kernel involving
$\smol_R^\dagger$ vanish, and there holds $\bs K(t)=\bs K^{TT}(t)$. (This is
also explicitly checked in a mode-coupling approximation.)

To describe slow dynamics arising from a coupling of translational modes,
we follow the standard procedure of MCT and rewrite the diffusion kernel
$\bs K^{TT}(t)$ in terms of a friction kernel. To do so, introduce a further
projector,
\begin{equation}
  \projP'=-\sum_{l_1l_2}\delrho_{l_1}(\vec q)\rangle
  \psi_{l_1l_2}(\vec q)\langle\delrho^*_{l_2}(\vec q)\smol^\dagger_T
\end{equation}
where $\bs\psi(\vec q)=\bs\omega_T^{-1}(\vec q)$ normalizes the projector.
We now decompose the propagator that appears in
Eq.~\eqref{eq:memk} according to
\begin{multline}
  e^{\projQ\smol^\dagger_T\projQ t}
  =e^{\projQ\smol^\dagger_T\projQ'\projQ t}
\\
  +\int_0^tdt'\,e^{\projQ\smol^\dagger_T\projQ(t-t')}\projQ\smol^\dagger_T\projP'\projQ
  e^{\projQ\smol^\dagger_T\projQ'\projQ t'}\,.
\end{multline}
This results in
\begin{equation}\label{eq:mk}
  \bs K^{TT}(\vec q,t)=\bs M(\vec q,t)-\int_0^t\bs K^{TT}(\vec q,t-t')\cdot
  \bs\omega_T^{-1}(\vec q)\cdot\bs M(\vec q,t')
\end{equation}
where we have defined the friction memory kernel
\begin{equation}\label{eq:memm}
  M_{ll'}(\vec q,t)=\left\langle\delrho_l^*(\vec q)\smol^\dagger_T\projQ
  e^{\projQ\smol^\dagger_T\projQ'\projQ t}\projQ\smol^\dagger_T\delrho_{l'}(\vec q)\right\rangle\,.
\end{equation}
Note that $\bs M(\vec q,t)$ and $\bs K^{TT}(\vec q,t)$ only differ
in their time evolution. In the context of passive Brownian particles,
the operator appearing in Eq.~\eqref{eq:memm} is also referred to as
the one-particle irreducible Smoluchowski operator \TODO{cite someone}.

Equations \eqref{eq:mz1} and \eqref{eq:mk} can be combined to a
time-evolution equation for the density correlation functions that is
a suitable starting point for approximations of the slow dynamics
arising from the slow evolution of positional density fluctuations,
\begin{subequations}\label{eq:mztwostep}
\begin{multline}\label{eq:mz2}
  \bs\omega_T^{-1}(\vec q)\cdot\partial_t\bs S(\vec q,t)
  +\left[\bs S^{-1}(q)+\bs\omega_T^{-1}(\vec q)\cdot\bs\omega_R\right]
  \cdot\bs S(\vec q,t)
  \\
  +\int_0^tdt'\,\bs m(\vec q,t-t')\cdot\left(
  \partial_{t'}\bs S(\vec q,t')
  +\bs\omega_R\cdot\bs S(\vec q,t')\right)
  =\bs0
\end{multline}
Here we have used that $\bs\omega_R\cdot\bs S^{-1}(q)=\bs\omega_R$ for
the spherical ABP system we consider, and abbreviated
\begin{equation}
\bs m(\vec q,t)=\bs\omega_T^{-1}(\vec q)\cdot\bs M(\vec q,t)
\cdot\bs\omega_T^{-1}(\vec q)\,.
\end{equation}
\end{subequations}

Equations \eqref{eq:mztwostep} are the starting point of mode-coupling
approximations for glassy dynamics. Setting $v_0=0$, the matrices
all become diagonal (since $\smol^\dagger_T$ does not mix translational
and rotational degrees of freedom in this case), and one recovers
for $S_{00}(q,t)$ the standard Mori-Zwanzig equation used to derive
MCT for Brownian spherical particles.

Rotational diffusion appears in Eqs.~\eqref{eq:mztwostep} in the form of
a ``hopping term'' in the MCT language, viz.\ the last term under the
integral in Eq.~\eqref{eq:mz2}. The original MCT only
contains a convolution of the memory kernel with the time derivative of
the density correlation function, viz.\ the first term under the integral in
Eq.~\eqref{eq:mz2}. In this form the equations allow for an
ideal glass transition: there exist solutions with a non-zero long-time
limit $\lim_{t\to\infty}\bs S(\vec q,t)=\bs F(\vec q)\neq\bs0$.
The presence of the density correlator itself
in the convolution integral for $l\neq0$ causes the corresponding
solutions to ultimately decay
exponentially: since rotation remains unhindered even in the dense
system, the associated density fluctuations will decay on a time scale
$1/(l^2D_r)$. Note that Eq.~\eqref{eq:mztwostep}
still allows for an ideal glass transition for all transient density
correlation functions $S_{ll'}(\vec q,t)$ with $l=0$.

The MCT approximation now consists of two intertwined steps: first the fluctuating
forces $\projQ\smol^\dagger_T\delrho_l(\vec q)$ that appear in $\bs M(\vec q,t)$
are replaced by their overlap with density-fluctuation pairs. Using the
short-hand notation $\delrho_1\equiv\delrho_{l_1}(\vec q_1)$, one introduces
the pair-density projector
\begin{equation}
  \projP_2=\sum_{1,2,1',2'}\delrho_1\delrho_2\rangle\chi_{121'2'}
  \langle\delrho_{1'}^*\delrho_{2'}^*
\end{equation}
with a suitable normalization matrix $\chi$.
Second, the resulting dynamical four-point correlation functions
that involve the reduced dynamics are replaced by the product of
two-point correlation functions propagated by the full dynamics,
\begin{multline}
  \langle\delrho_1^*\delrho_2^* e^{\projQ\smol^\dagger_T
  \projQ'\projQ t} \delrho_{1'}\delrho_{2'}\rangle
\\
  \approx \langle\delrho_1^* e^{\adsmol t} \delrho_{1'}\rangle
  \langle\delrho_2^* e^{\adsmol t} \delrho_{2'}\rangle
  +\{1'\leftrightarrow 2'\}
\end{multline}
together with a consistent approximation of $\chi$.
For a detailed derivation of the MCT expression for the memory kernel,
we refer to Appendix~\ref{sec:mctderiv}.
One gets
\begin{multline}
  m_{ll'}(\vec q,t)=\frac{\rho^2}{2N}\sum_{\vec k+\vec p=\vec q}
  \sum_{l_1l_2l_1'l_2'}
  \mathcal V^\dagger_{ll_1l_2}(\vec q,\vec k,\vec p)
  \times \\ \times 
  S_{l_1l_1'}(\vec k,t)S_{l_2l_2'}(\vec p,t)
  \mathcal V_{l'l_1'l_2'}(\vec q,\vec k,\vec p)
\end{multline}
The vertices are given by $\mathcal V^\dagger_{ll_1l_2}(\vec q,\vec k,\vec p)
=\sum_m(\bs\omega_T^{-1}(\vec q))_{lm}\mathcal W^\dagger_{ml_1l_2}(\vec q,\vec k,\vec p)$ and
$\mathcal V_{ll_1l_2}(\vec q,\vec k,\vec p)=\sum_m(\bs\omega_T^{-1})_{lm}(\vec q)\mathcal W_{ml_1l_2}(\vec q,\vec k,\vec p)$, with
\begin{multline}
  \mathcal W^\dagger_{ll_1l_2}(\vec q,\vec k,\vec p)=D_t\delta_{l,l_1+l_2}
  \left(\vec q\cdot\vec k\,c_{l_1l_1}(k)+\vec q\cdot\vec p\,c_{l_2l_2}(p)\right)\\
  +\frac{iv_0}{2\rho}\delta_{|l-l_1-l_2|,1}S_{l_1l_1}(q)
  \Bigl(ke^{-i(l-l_1-l_2)\theta_k}\tilde S_{l-l_2,l_1}(k)
\\
  +pe^{-i(l-l_1-l_2)\theta_p}\tilde S_{l-l_1,l_2}(p)
  -qe^{-i(l-l_1-l_2)\theta_q}\Bigr)
\end{multline}
where we have defined
$\tilde S_{ll'}(k)=S_{l'l'}^{-1}(k)S_{l,l}(k)$.
The left vertex is the same as in equilibrium,
\begin{equation}
  \mathcal W_{ll_1l_2}(\vec q,\vec k,\vec p)
  =D_t\delta_{l,l_1+l_2}\left(\vec q\cdot\vec k\,c_{l_1l_1}(k)
  +\vec q\cdot\vec p\,c_{l_2l_2}(p)\right)
\end{equation}
It differs from $\mathcal W^\dagger$ because the time-evolution operator
$\adsmol$ is not self-adjoint with respect to the scalar product defined
by the equilibrium averages.

For $v_0=0$, these equations reduce to the equilibrium MCT expressions.
In particular, in this case $\mathcal V$ and $\mathcal V^\dagger$ do
no longer explicitly depend on $D_t$.
In the active case, $v_0\neq0$, one readily checks that the explicit dependence
on $D_t$ still cancels in the memory kernel if the self-propulsion
velocity is expressed in terms of a P\'eclet number
$\tlname{Pe}_t=v_0\sigma/D_t$, where $\sigma$ is a typical particle
diameter.
There is no dependence on $D_r$ in the memory kernel, and therefore
the P\'eclet number $\tlname{Pe}=v_0^2/D_tD_r$ does not assume the natural
role in determining the MCT dynamics that is has for the low-density
system.

The MCT approximation preserves the transformation properties of
the correlation functions under rotation, Eq.~\eqref{eq:rotation}.
In fact, the same transformation law is required for all the quantities that
appear in the
Mori-Zwanzig equations, Eq.~\eqref{eq:mz1} or \eqref{eq:mz2}. In particular,
$\bs\omega(\vec q)\mapsto\bs u(\alpha)\cdot\bs\omega(\vec q^{\,\prime}\cdot
\bs\omega^\dagger(\alpha)$ is easily checked.
For the MCT vertices, a straightforward calculation shows
$\mathcal W_{ll_1l_2}^\dagger(\vec q,\vec k,\vec p)
\mapsto\mathcal W_{l'l_1'l_2'}^\dagger(\vec q^{\,\prime},\vec k^{\,\prime},
\vec p^{\,\prime})u_{ll'}(\alpha)u^\dagger_{l_1'l_1}(\alpha)u^\dagger_{l_2'l_2}(\alpha)$ and equivalently for $\mathcal W_{ll_1l_2}(\vec q,\vec k,\vec p)$,
i.e., all terms that appear in the MCT expression for $\bs m(\vec q,t)$
transform like tensors.

The fact that the MCT approximation preserves the transformation properties of
the correlation functions under rotation, allows us to pick
a coordinate system where $\vec q$ is aligned along a coordinate
axis, $\vec q=q\vec e_y$, say. By using the unitary transformation
property, all correlation functions entering the MCT memory kernel
can be rewritten in terms of those evaluated with $\vec q$ aligned
along the same axis. For completeneess, this form of the MCT equations
is documented in Appendix~\ref{sec:mctnum}.
This allows to reduce the numerical calculation
to wave vectors along a single spatial axis.

The MCT equations have been solved numerically on a grid of \TODO{check} {$128$}
wave numbers equally spaced up to \TODO{check} {$|\vec q|\le Q$ with $Q\sigma=50$}.
For the angular indices, a cutoff $|l|\le L$ with $L=1$ has been introduced.
Some results have been checked with $L=2$, but the effects were minor.
The presence of the ``hopping'' term with its singular structure (imposed
by $\omega_{R,00}=0$) poses a numerical problem at long times.
We have developed an extension of the standard algorithm that is usually
employed to solve MCT equations, and that we outline in
Appendix~\ref{sec:mctalgo}.

To determine the MCT vertex, the equilibrium static structure factor is
needed. In odd dimensions, the Percus-Yevick approximation provides a
reasonably accurate analytical expression. However, in even dimensions,
no such analytical solution is known. We use the Baus-Colot expression
for $S(q)$ that is known to be close to simulation data.

We fix units of length and time by the particle diameter $\sigma=1$ and
the translational free-diffusion time $\sigma^2/D_t=1$.
Densities are reported as packing fractions, $\phi=(\pi/4)\rho\sigma^2$.
With the chosen parameters, we obtain a glass transition in the passive
hard-disk system at a critical packing fraction $\phi_c\approx0.7207$;
\TODO{check and update}
this is in reasonable agreement with the value \TODO{xxx} reported in
earlier work \cite{glass-theory:Bayer.2007}.
In comparison to this work, we have improved the numerical evaluation
of the wave-vector integrals appearing in the MCT memory kernel; see
Appendix~\ref{sec:mctnum} for details.

\section{Dynamics}\label{sec:dynamics}

\begin{figure}
\includegraphics[width=\linewidth]{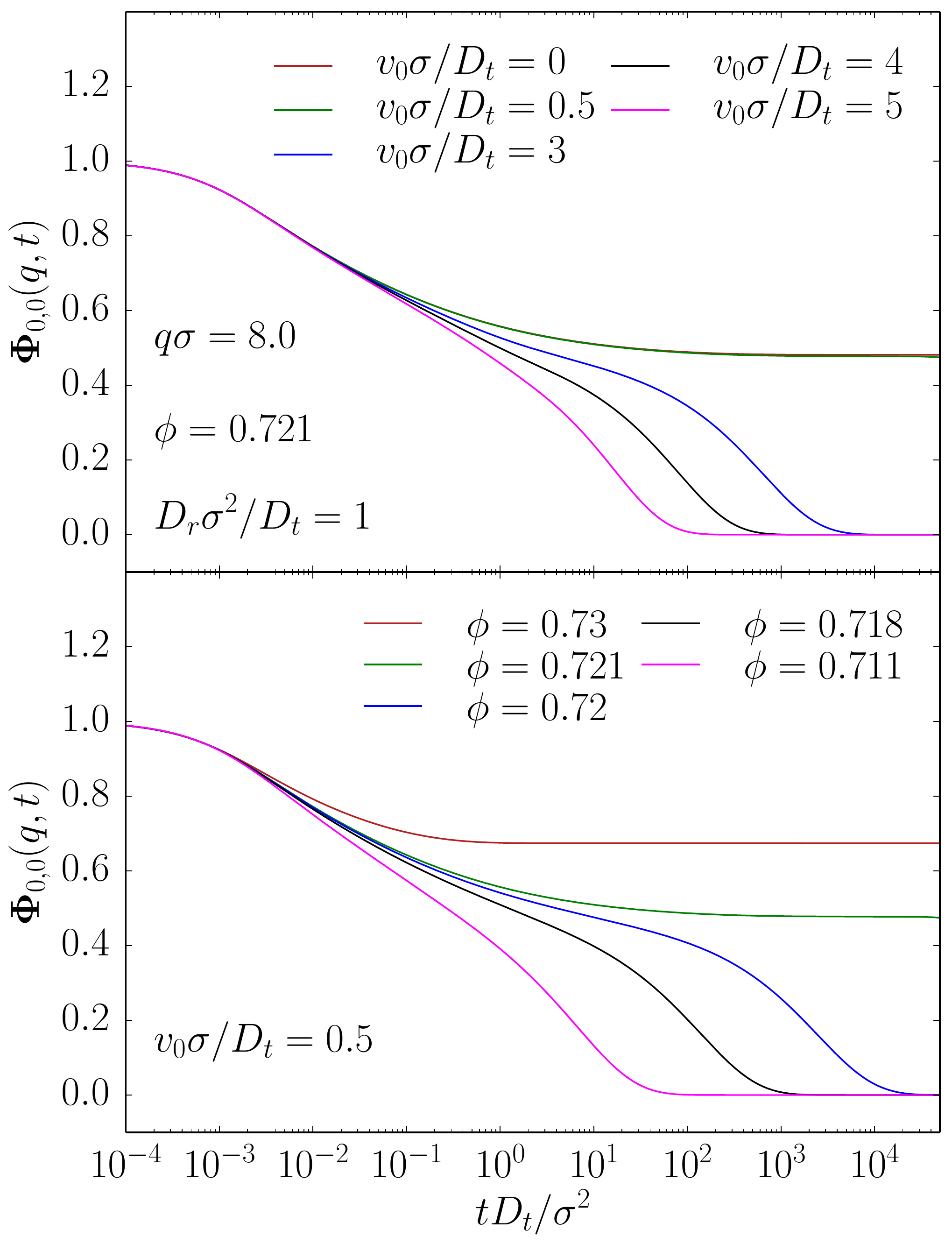}
\caption{\label{fig:correlators}
  Transient density correlation functions $\Phi_{00}(q,t)$ of a hard-disk ABP
  system within MCT, at wave number $q\sigma=8$ and for $D_r=1$.
  Upper panel: functions at constant packing fraction $\phi=0.721$ above
  the passive glass transition, for increasing
  $v_0$ (right to left) as labeled.
  Lower panel: functions at constant self-propulsion velocity $v_0$,
  for increasing packing fraction (left to right).
}
\end{figure}

Exemplary MCT results for the density correlation functions
$\Phi_{00}(\vec q,t)$ are shown in Fig.~\ref{fig:correlators}. A wave number
$q\sigma=8$ in the vicinity of the main peak of the equilibrium $S(q)$ was
chosen.
The density correlation functions show the qualitative features expected
for dense colloidal suspensions nearing dynamical arrest: after an initial
relaxation, a plateau emerges at high densities that extends over an
increasingly large intermediate-time window as the density is increased.
The final relaxation from this plateau to zero is termed structural
relaxation, and its characteristic time increases strongly with increasing
density. At the highest densities shown, structural relaxation becomes
ineffective and is not seen over the full time window accessible to the
numerical solution scheme. Hence, $\lim_{t\to\infty}\Phi_{00}(\vec q,t)
=F_{00}(\vec q)>0$. This non-zero positive non-ergodicity
parameter signals the appearance of an ideal glass.

With increasing self-propulsion velocity $v_0$, the dynamics speeds up,
as shown in the upper panel of the figure. This is qualitatively expected,
since self-propulsion renders the particle motion more vivid, and this
opposes the slow dynamics. This result is also qualitatively consistent
with earlier Brownian-dynamics (BD) simulation studies of three-dimensional
ABP \TODO{whom to cite} \cite{biophys:Ni.2013}. (Note that there, stationary-state
non-equilibrium correlation functions were reported, while the central object
of our MCT are the transient non-equilibrium, equilibrium-averaged correlation
functions.)

For small enough self-propulsion velocity, the glass remains stable at
sufficiently high density. This
is exemplified by the lower panel of
Fig.~\ref{fig:correlators}, where curves for constant $v_0=0.5\,D_t/\sigma$
are shown. At the highest packing fractions shown, no sign
of structural relaxation is seen in the numerical results over the
time window covered in the figure.

Hence, for small enough but finite $v_0$ MCT predicts an ideal ``active glass''.
As the lower panel of Fig.~\ref{fig:correlators} demonstrates, the signature
of the transition to this active glass is qualitatively as for the
passive ideal glass: with increasing density at fixed $v_0$, structural
relaxation dramatically slows down until it completely arrests at the
transition density $\phi_c$.
Further increasing the density causes the nonergodicity
parameter to increase. The ideal glass transition is discontinuous in the
sense that the long-time limit of the density correlation functions
jumps from zero in the liquid ($\phi<\phi_c$) to a finite value
at $\phi_c$.

These results suggest that there is a line of ideal glass transitions
$(\phi_c,v_0^c)$ in the density--self-propulsion plane. This line
shifts to increasing density with increasing $v_0$.
Qualitatively, this result has been derived in earlier extensions of MCT
that do not account for orientational degrees of freedom explicitly
\cite{biophys:Farage.2014pre}.

\begin{figure}
\includegraphics[width=\linewidth]{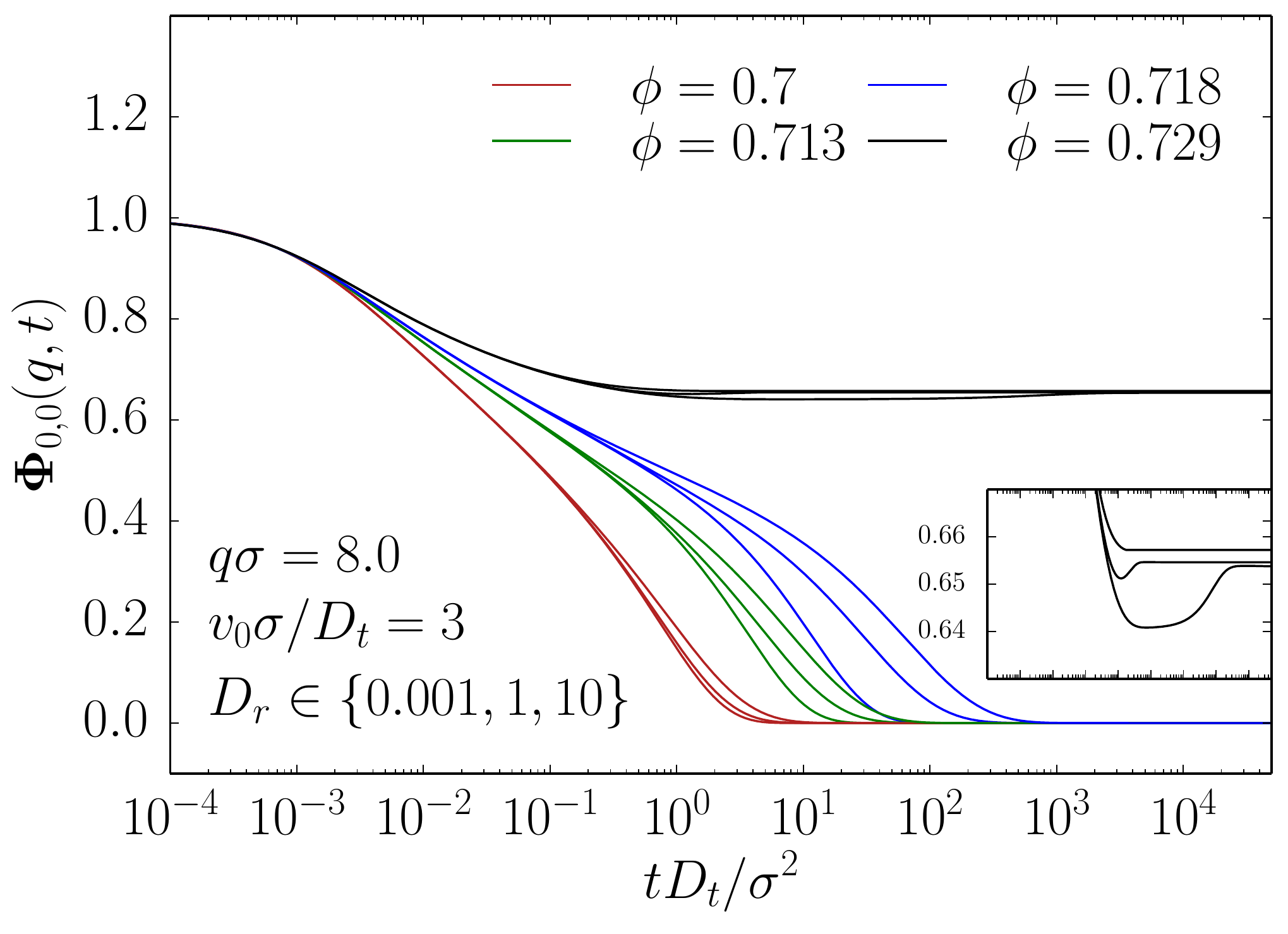}
\caption{\label{fig:correlators_dr}
  Density correlation functions $\Phi_{00}(\vec q,t)$ for packing
  fractions $\phi=0.7$, $0.713$, $0.718$, and $0.729$ (groups of lines
  from left to right)
  and self-propulsion velocity $v_0=3\,D_t/\sigma$, for different values of the
  rotational diffusion coefficient, $D_r=1/1000$, $1$, and $10$ (left to
  right).
  The inset provides a zoom of the $\phi=0.729$ curves.
}
\end{figure}

With the present approach, the influence of
the rotational diffusion coefficent on the dynamics can be studied.
Figure~\ref{fig:correlators_dr} exemplifies the effects of increasing
$D_r$ on the dynamics at fixed self-propulsion velocity.
Keeping the other parameters fixed, an increase in $D_r$ leads to a
slowing down of the structural-relaxation dynamics.
Qualitatively, this is expected from the argument that faster reorientation
of the individual particles causes the self propulsion to be less effective
in melting nearest neighbor cages, since for the latter process a certain
persistence of the self-propulsion force in a specific direction needs to
be maintained.
The slowing down with increasing $D_r$ is more pronounced at higher densities:
while at $\phi=0.7$, the final relaxation of the curves shown
in Fig.~\ref{fig:correlators_dr} spreads out by about a factor $2$,
the same change in $D_r$ causes the structural-relaxation time to change
by a factor of about $10$ at $\phi=0.718$.

\begin{figure}
\includegraphics[width=\linewidth]{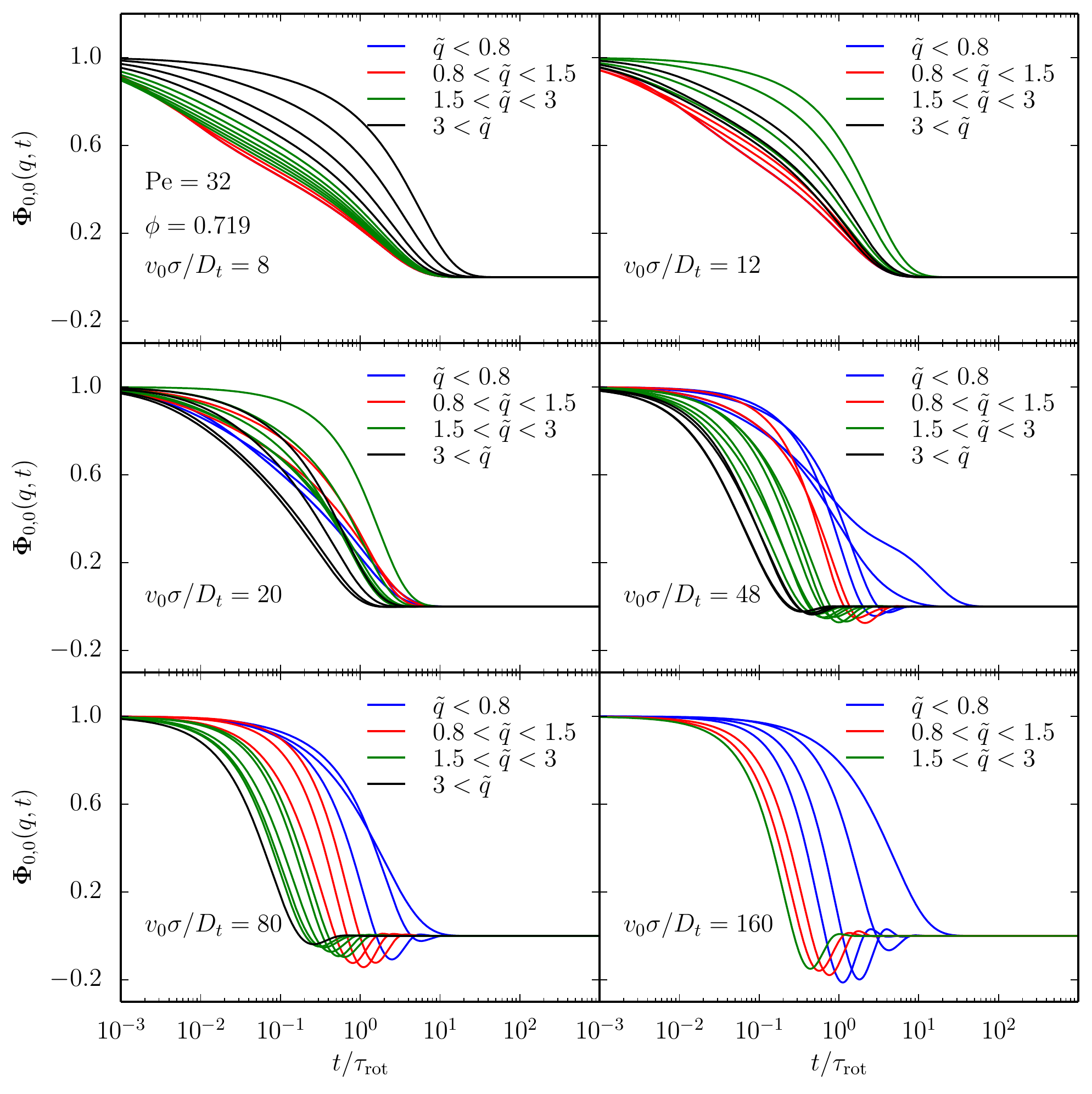}
\caption{\label{fig:correlators_q}
  Wave-number dependence of the transient
  density correlation functions $\Phi_{00}(q,t)$ for packing fraction
  $\varphi=0.719$ just below the passive-glass transition,
  for different velocities as labeled, keeping the
  P\'eclet number $\tlname{Pe}=v_0^2/D_rD_t=32$ fixed.
  From top left to bottom right, the values are
  $(v_0,D_r)=(8,2)$, $(12,4.5)$, $(20,12.5)$, $(48,64)$, $(80,200)$,
  and $(160,800)$.
  Wave numbers $\tilde q=q\ell_p/(2\pi)$ are given in units of the persistene
  length, $\ell_p=v_0/D_r$, and increase from right to left (in the top
  part of the curves). The panels correspond to
  $\ell_p=4$, $2.67$, $1.6$, $0.75$, $0.4$, and $0.2$.
  Correlators are shown as functions of rescaled time, $t/\tau_\text{rot}$,
  with the time scale set by the reorientational diffusion,
  $\tau_r=\sigma/D_r$.
}
\end{figure}

The structural relaxation dynamics of the weakly active system shows the
same qualitative features as they are known from the passive system. In
particular, the structural relaxation process can be well described by
a stretched-exponential function in time for the cases shown
in Figs.~\ref{fig:correlators} and \ref{fig:correlators_dr}.
It is known from the passive system, that density fluctuations with
wavelengths comparable to the particle size govern the slow dynamics.

For the active system, the persistence of self propulsion sets another length
scale, $\ell_p=v_0/D_r$.
This suggests to discuss the effect of density fluctuations on length
scales much larger, comparable to, and much smaller than the persistence
length.
Let us introduce a rescaled wave number $\tilde q=q\ell_p/(2\pi)$.
The low-density dynamics of ABP exhibits three distinct regimes
\cite{biophys:Kurzthaler.2016}: for $\tilde q\ll1$ probe the dynamics on
length scales large compared to $\ell_p$, and hence see diffusive
relaxation with a diffusion coefficient $D_\text{eff}$. For small length
scales, probed by $\tilde q\gg1$, the initial Brownian passive diffusion
of the ABP is seen, and the density correlators decay diffusively, with
diffusion coefficient $D_t$. Activity causes an intermediate regime
$\tilde q\approx1$ to appear, where the persistent swimming motion affects
the relaxation of density fluctuations. Over the length scales probed in
this regime, particles swim in a fixed direction and cause density fluctuations
to decay in a damped-oscillatory fashion, leading to pronounced ``undershoots''
in the final relaxation.

The influence of high-density interactions on this single-particle picture
is examined in Fig.~\ref{fig:correlators_q}.
Here, state points close to the glass transition were chosen and to
make connection to the low-density theory, different activities along a
cut with constant P\'eclet number, $\tlname{Pe}=32$, are shown.
The panels of Fig.~\ref{fig:correlators_q} correspond to increasing
self-propulsion velocity (from top left to bottom right), and
at fixed $\tlname{Pe}$, these correspond to decreasing persistence lengths.
Essentially, the low-density scenario is recovered for the case $\ell_p<\sigma$,
i.e., for large self-propulsion velocity. There, structural arrest is
effectively destroyed by active driving, and the correlation functions decay
on a time scale $\tau_\text{rot}=1/D_r$, showing oscillations around
$\tilde q\approx1$ (cf.\ lower panels of Fig.~\ref{fig:correlators_q}).

For large persistence length, $\ell_p\gg\sigma$, the top panels of
Fig.~\ref{fig:correlators_q} demonstrate that the low-density scenario is
absent. Here, all correlation functions decay without oscillations.
This result can be interpreted as showing that once the interparticle
length scale becomes smaller than $\ell_p$, density fluctuations can no
longer be translated by persistent motion; they relax by the combination of
diffusion and activity-modified structural interactions.
Coincidently, for the $\ell_p\gg\sigma$ cases shown in Fig.~\ref{fig:correlators_q},
the intermediate-time plateau
of structural relaxation begins to emerge at large $\tilde q$ (corresponding
to intermediate $q\sigma$).

\TODO{can we compare the curves in the lower panels with the analytical
low-$\varphi$ solution?}

The appearance of oscillatory relaxation in the density correlation functions
of a Brownian system is a clear signature of non-equilibrium dynamics.
Recall that the equilibrium Smoluchowski operator $\smol_\text{eq}$ is
negative semi-definite, i.e., it has non-positive real eigenvalues only.
As a result, the corresponding auto-correlation functions are completely
monotone functions: they can be written as superpositions of purely relaxing
exponential functions with positive weights.
This is a feature that is preserved under the MCT approximation.

\TODO{Farage Brader, their extension, and Szamel/Berthier AOUP: I think
that these MCT equations are of the form that they preserve complete
monotonicity? check: In this context, it is not important that the
operator violates
the FDT, because still a stationary solution to vanishing current can be
constructed, i.e., an equilibrium pdf?? From this, check that one can show
that all matrix elements of $\smol_\text{eff}$ are negative semi-definite.}

\begin{figure}
\includegraphics[width=\linewidth]{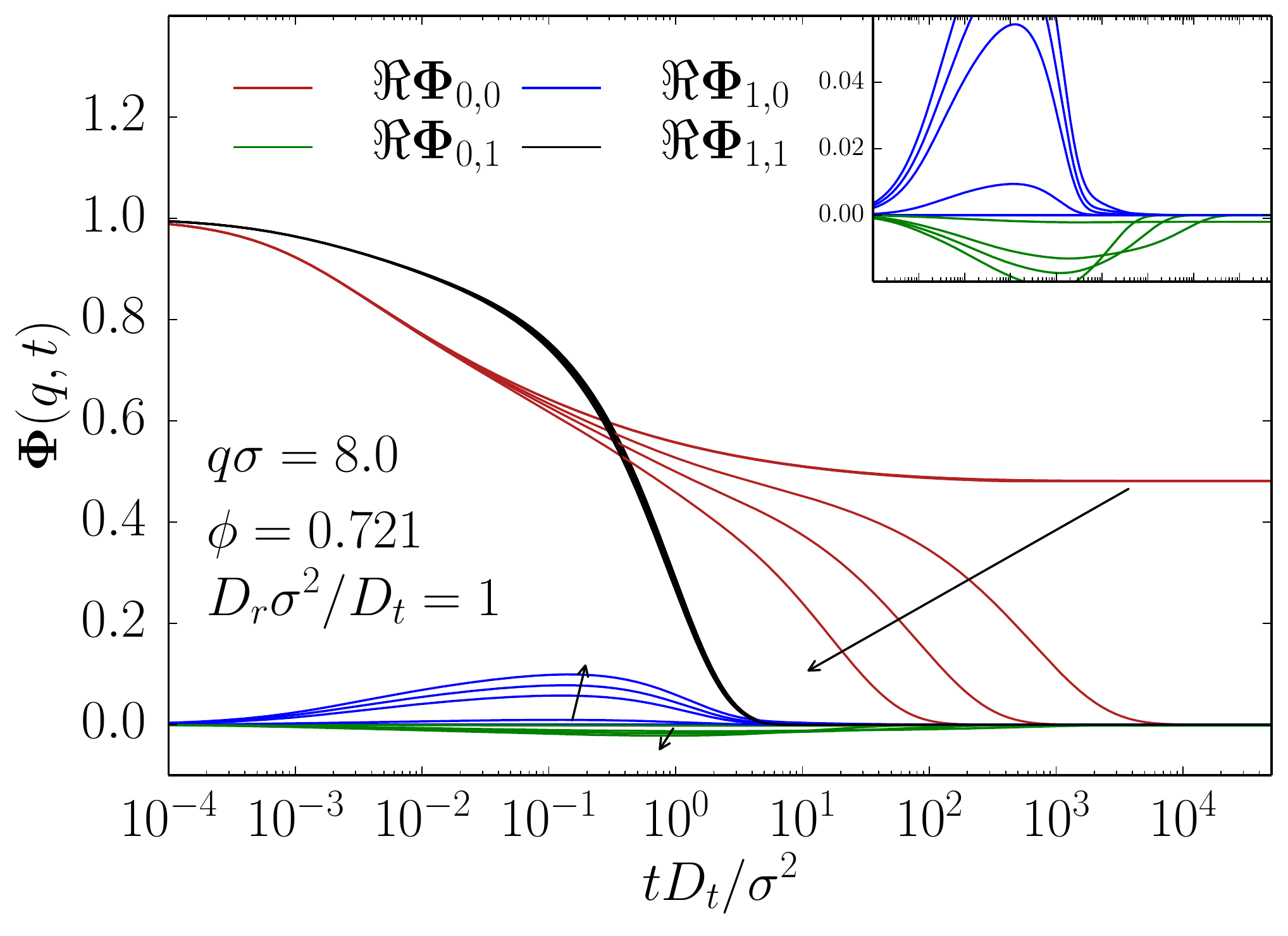}
\caption{\label{fig:correlators_l}
Matrix elements of the transient density correlation matrix $\bs\Phi(\vec q,t)$
at fixed packing fraction $\phi$ and rotational diffusion $D_{r}$,
for increasing $v_0$ (right to left).
The inset provides a zoom for the off-diagonal components.
}
\end{figure}

To highlight the dynamics of the reorientational degrees of freedom,
we show in Fig.~\ref{fig:correlators_l} the matrix elements of
$\Phi_{ll'}(\vec q,t)$ for $|l|\le1$.
Only positive $l,l'$ are shown for siplicity, and the case $l=l'=0$
is repeated from above for reference.
The $(ll')=(11)$ correlator reflects the decay of orientational order.
It decays on the time scale $\tau_\text{rot}=\sigma/D_r$,
with a final exponential relaxation. This is expected since in the
spherical ABP model, rotation is not influenced by the packing of particles.
Note that the decay of $\Phi_{11}(\vec q,t)$ is not purely exponential;
even at low densities and without self propulsion,
$\Phi_{11}(\vec q,t)\sim\exp[-q^2D_tt]\exp[-D_rt]$.
In general, the $(ll')=(11)$ correlator inherits a signature of the
translational motion for $t\ll\tau_\text{rot}$, which is cut off
by an exponential decay at $t\sim\tau_\text{rot}$.

The off-diagonal elements $(ll')=(01)$ and $(10)$ of the transient
density correlation function behave different from each other.
In equilibrium, one expects the matrix of correlation functions to be
symmetric (as is the case for example in the MCT developed for Newtonian
non-spherical particles \cite{glass-theory:Franosch.1997c,glass-theory:Kaemmerer.1997,glass-theory:Schilling.2002}).
This symmetry is lost in the present theory because the
time-evolution operator is not self-adjoint with respect to the
equilibrium-weighted scalar product.

The off-diagonal elements vanish with $v_0\to0$, as expected from the
structure of the Smoluchowski equation for spherical ABP. Interestingly,
the correlation function $\Phi_{01}(\vec q,t)$ shows non-trivial slow
dynamics that is coupled to the slow dynamics of the positional-density
correlator $\Phi_{00}(\vec q,t)$. In particular it displays structural
relaxation that slows down beyond the time scale $\tau_r$ over which
orientational order decays. The correlation function $\Phi_{10}(\vec q,t)$
instead decays on the reorientational time scale $\tau_\text{rot}$.

This observation can be rationalized by the peculiar structure of the
equations of motion of spherical ABP: orientations influence the slow dynamics
of the positions, but not vice versa. Thus, the evolution of positional
density fluctuations, $\exp[\smol^\dagger t]\delrho_0(\vec q)$, will not
be detectable in the subspace spanned by the orientations,
$\delrho_1^*(\vec q)$, for times $t\gg\tau_\text{rot}$. On the other hand,
the impact of the time-evolved initial polar order,
$\exp[\smol^\dagger t]\delrho_1(\vec q)$, on the positional density
fluctuations of the system, $\delrho_0^*(\vec q)$, persists until the
time of overall structural relaxation, hence
$S_{01}(\vec q,t)=\langle\delrho_0^*(\vec q)\exp[\smol^\dagger t]\delrho_1(\vec q)\rangle$ decays as slow as $S_{00}(\vec q,t)$.

\begin{figure}
\includegraphics[width=\linewidth]{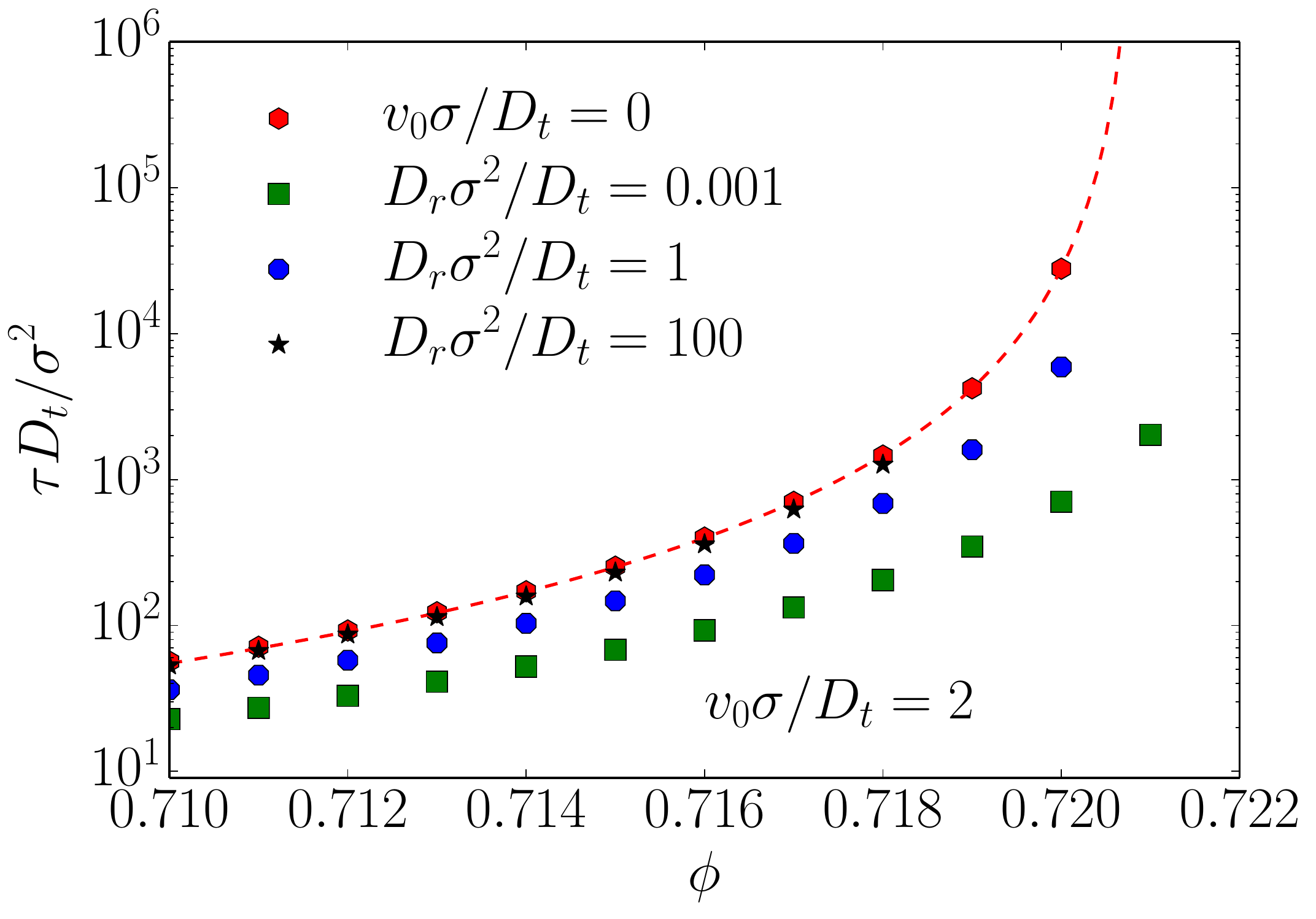}
\caption{\label{fig:tau}
  Structural relaxation times $\tau(\phi,v_0,D_r)$ as a function of $\phi$
  for fixed $v_0$ and different $D_r$ (as indicated by the diferent symbols).
  Hexagon symbols connected by a dashed
  line correspond to the passive system.
}
\end{figure}

The usual quantification of the slow dynamics is in terms of the
structural relaxation time $\tau$. Following the operational defition
used in many studies of glassy dynamics, we define $\tau$ as the time
where the density-correlation function has decayed to $1\%$ of its
initial value, $\Phi_{00}(\vec q,\tau)=0.01$. \TODO{check definition here and in the following figure}

Results for $\tau(\varphi)$ for a fixed self-propulsion velocity
are shown in Fig.~\ref{fig:tau}.
As anticipated from the discussion above, the structural relaxation time
strongly increases with increasing packing fraction, in a power-law
fashion that is the hallmark of the approach to the MCT glass transition.
At fixed $v_0$, increasing $D_r$ increases the structural relaxation time.
For $D_r\to\infty$, the $\tau$-vs-$\varphi$ curve corresponding to the
passive system is approached.

The data shown in Fig.~\ref{fig:tau} agree qualitatively with corresponding
3D results from BD simulations \cite{biophys:Ni.2013}. Again, similar results
have been discussed for the AOUP system \cite{biophys:Flenner.2016pre} and glassy tissue
models \cite{biophys:Bi.2016}, demonstrating that the MCT power law that describes the
increase in relaxation time is quite robust.

\begin{figure}
\includegraphics[width=\linewidth]{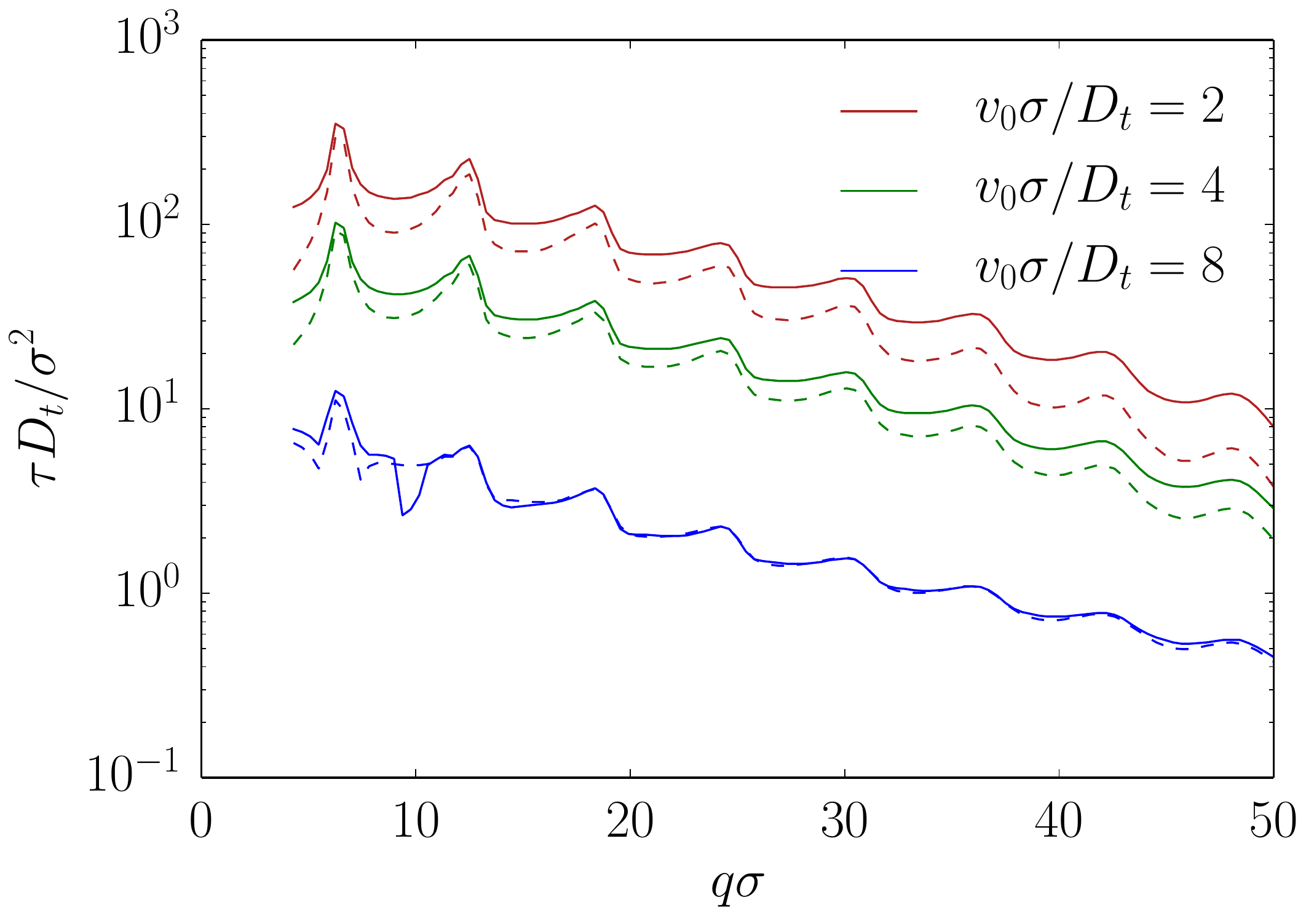}
\caption{\label{fig:tau_q}
Structural relaxation times $\tau(\phi,v_0,D_r)$ as a function of wavenumber $q$
for fixed $D_r=1\,D_t/\sigma^2$ and various $v_0$ as indicated (increasing
$v_0$ from top to bottom).
Solid lines are extracted from the positional-density correlator
$\Phi_{00}(\vec q,t)$, dashed lines from the orientational correlator
$\Phi_{01}(\vec q,t)$.
}
\end{figure}

The structural relaxation time depends on the wave number of the density
fluctuations. The $q$-dependence of $\tau$ is shown for a fixed density
and various self-propulsion velocities in Fig.~\ref{fig:tau_q}.
Both the slow relaxation times for $(ll')=(00)$ and $(01)$ are shown.
From the passive case, it is known that $\tau(q)$ is an oscillating
decaying function of $q$, with oscillations in phase with those of the
equilibrium static structure factor $S(q)$. This typical signature of
glassy dynamics is visible in Fig.~\ref{fig:tau_q} also for the active
system. Increasing self-propulsion velocity shifts the relaxation times
to shorter values, essentially by the same amount for all $q$ for the
range of $v_0$ shown. This
emphasizes that the active enhancement of structural relaxation in
this case is a collective effect. Note that here, $\ell_p>\sigma$,
i.e., the results are for the regime in Fig.~\ref{fig:correlators_q} where
density fluctuations cannot be shifted by persistent motion.

The relaxation times of the $(ll')=(01)$ correlator (dashed lines in
Fig.~\ref{fig:tau_q}) show the same qualitative behavior as those of the
$(ll')=(00)$ correlator. Approaching the glass transition, all slow
relaxation modes that are relevant within MCT become strongly coupled,
so that one expects the approach to a common scaling behavior. This is
indeed seen for $v_0=4\,D_t/\sigma$ and $v_0=2\,D_t/\sigma$ in the figure.
For the larger $v_0=8\,D_t/\sigma$, the $\tau$-versus-$q$ curves corresponding
to the two correlators ($l'=0$ and $l'=1$) become identical at large $q$.
At the same tie, the oscillations in $q$ become slightly less pronounced.
This indicates that the dynamics on short length scales and for strong
self propulsion looses its collective character and becomes more
incoherent.

\section{Glass Transition}\label{sec:glasstransition}

\begin{figure}
\includegraphics[width=\linewidth]{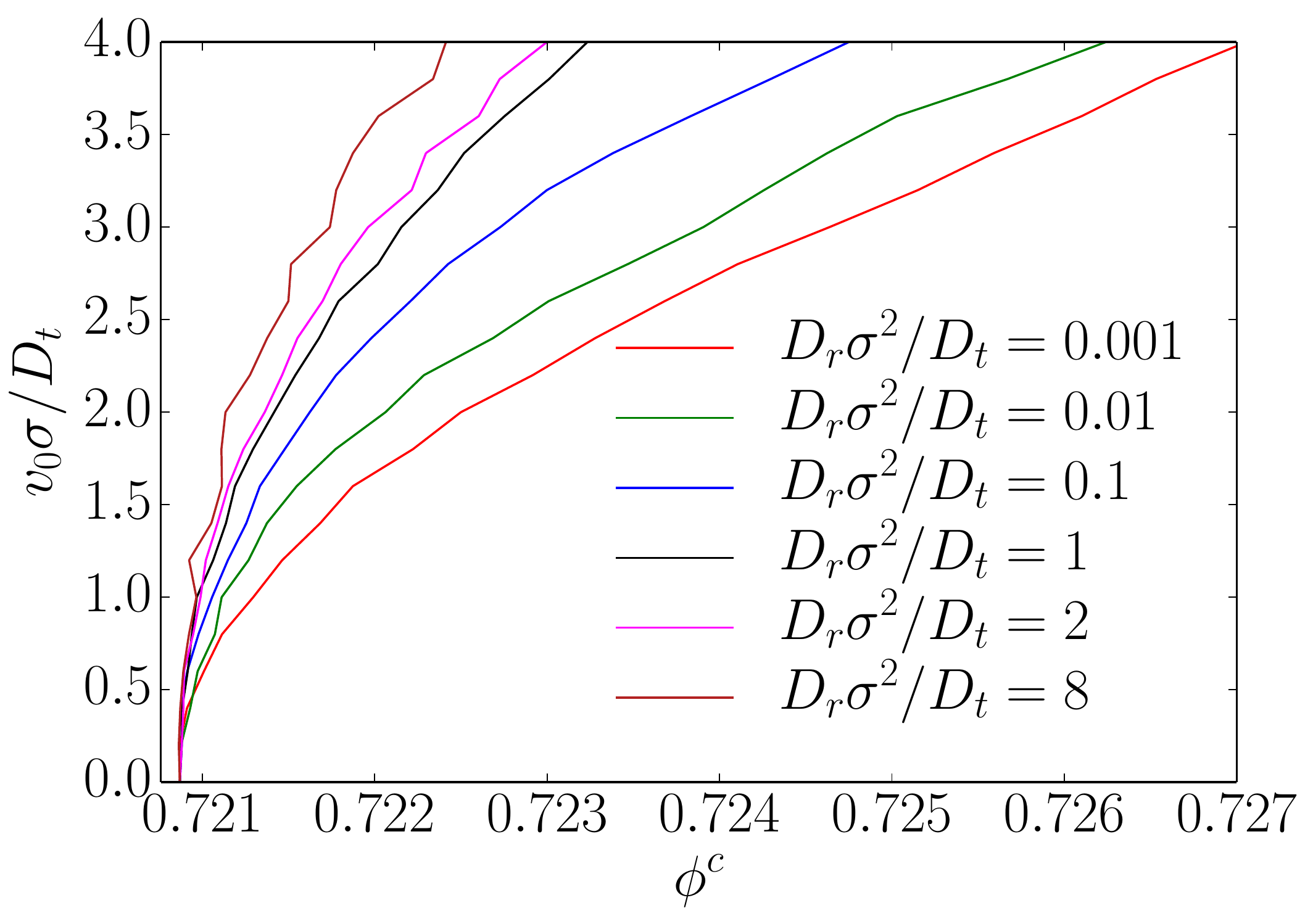}
\caption{\label{fig:glass_diagram}
  Active-glass-transition lines obtained from power-law fits to the
  structural relaxation times $\tau(\phi)$ obtained from the transient
  density correlators $\Phi_{00}(\vec q,t)$. Curves correspond to
  different choices for $D_r$ as labeled (increasing from bottom to top).
}
\end{figure}

The ideal glass transition is signalled by the appearance of a non-zero
nonergodicity parameter, $\bs F(\vec q)=\lim_{t\to\infty}\bs S(\vec q,t)
\neq\bs0$.
In the standard MCT, one derives a separate algebraic equation for
$\bs F(\vec q)$ from the long-time behavior of the equations of motion.
This assumes that the solutions $\bs S(\vec q,t)$ are slowly varying
functions, such that the time derivatives of the correlation functions
become arbitrarily small at long times.
The same procedure applies here in the case $D_r=0$, where $\bs\omega_R=\bs0$.
Then, one arrives at
\begin{equation}\label{eq:mf0}
\bs F(\vec q)+\bs m(\vec q)\cdot(\bs F(\vec q)-\bs S(q))=\bs0
\end{equation}
where we use the short-hand $\bs m(\vec q)\equiv\lim_{t\to\infty}\bs m(\vec q,t)$.
Equation~\eqref{eq:mf0} is a nonlinear implicit equation for $\bs F(\vec q)$,
since $\bs m(\vec q)$ is a bilinear functional of these matrices.
Generically, there appear bifurcation points where the physically relevant
solution of Eq.~\eqref{eq:mf0} changes from $\bs F(\vec q)=\bs0$ to
some $\bs F(\vec q)\neq\bs0$. These bifurcations indicate idealized
glass-transition points \cite{glass-theory:Goetze.2009}.
Note that Eq.~\eqref{eq:mf0} is an algebraic equation that can be evaluated
without solving the time-dependent MCT equations.

The presence of the term $\bs\omega_R\neq\bs0$ complicates the determineation
of the $t\to\infty$ limit of $\bs S(\vec q,t)$. From a Laplace transform
of Eq.~\eqref{eq:mztwostep}, one finds that the long-time limits need to
obey
\begin{subequations}
\begin{equation}
  \bs m(\vec q)\cdot\bs\omega_R\cdot\bs F(\vec q)=\bs 0\,,
\end{equation}
together with
\begin{equation}\label{eq:mf1}
  \bs F(\vec q)+\bs m(\vec )\cdot(\bs F(\vec q)-\bs S(q)
  +\bs\omega_R\cdot\bs S_0(\vec q))=\bs 0\,,
\end{equation}
\end{subequations}
where $\bs\omega_R\bs S_0(\vec q)=\int_0^\infty dt\,\bs\omega_R\cdot\bs S(\vec q,t)$ is the integral over the decaying matrix elements of the correlator.
Here, we have assumed that the ultimate relaxation to the long-time
value is faster than algebraic, motivated by the presence of exponentially
decaying $l\neq0$ modes. Unlike Eq.~\eqref{eq:mf0}, Eq.~\eqref{eq:mf1} is
no longer an equation that involves only the long-time limits of the
correlators and their memory kernels; through $\bs S_0(\vec q)$, details
on the full time evolution enter.
Still, bifurcation points in Eq.~\eqref{eq:mf1} should signal idealized
glass transitions, and the asymptotic analysis of the MCT equations close
to these transition points proceeds in analogy to the passive case
\cite{glass-theory:Goetze.2009}. However, with our current algorithm we found
the numerical evaluation of the bifurcation points of Eq.~\eqref{eq:mf1}
to be too unstable,
because they depend sensitively on a precise evaluation of $\bs S_0(\vec q)$.

We have
therefore determined tentative glass-transition points from
extrapolations of the $\tau$-vs-$\phi$ curves using the expected
asymptotic MCT power laws, $\tau\sim|\phi-\phi_c|^{-\gamma}$.
All three parameters (amplitude, exponent and critical point) in this
asymptotic formula were allowed to depend on the model parameters
$v_0$ and $D_r$.
The results for various fixed $D_r$ are shown in Fig.~\ref{fig:glass_diagram}
as glass-transition lines in the $(\phi,v_0)$ plane.
The transition lines depend on both $v_0$ and $D_r$ explicitly, and we
did not observe a collapse of the curves when either $\tlname{Pe}$ or
$\ell_p$ are kept fixed.

Qualitatively, these extrapolations confirm the observations made above:
increasing self-propulsion speed shifts the glass transition to higher
densities, and increasing the rotational diffusion coefficient
shifts the transition to larger $v_0$. In particular, as $D_r\to\infty$,
the glass-transition lines approach a vertial line in the $(\phi,v_0)$ plane,
i.e., become independent of $v_0$ and identical to the passive $\phi^c$.

\begin{figure}
\includegraphics[width=\linewidth]{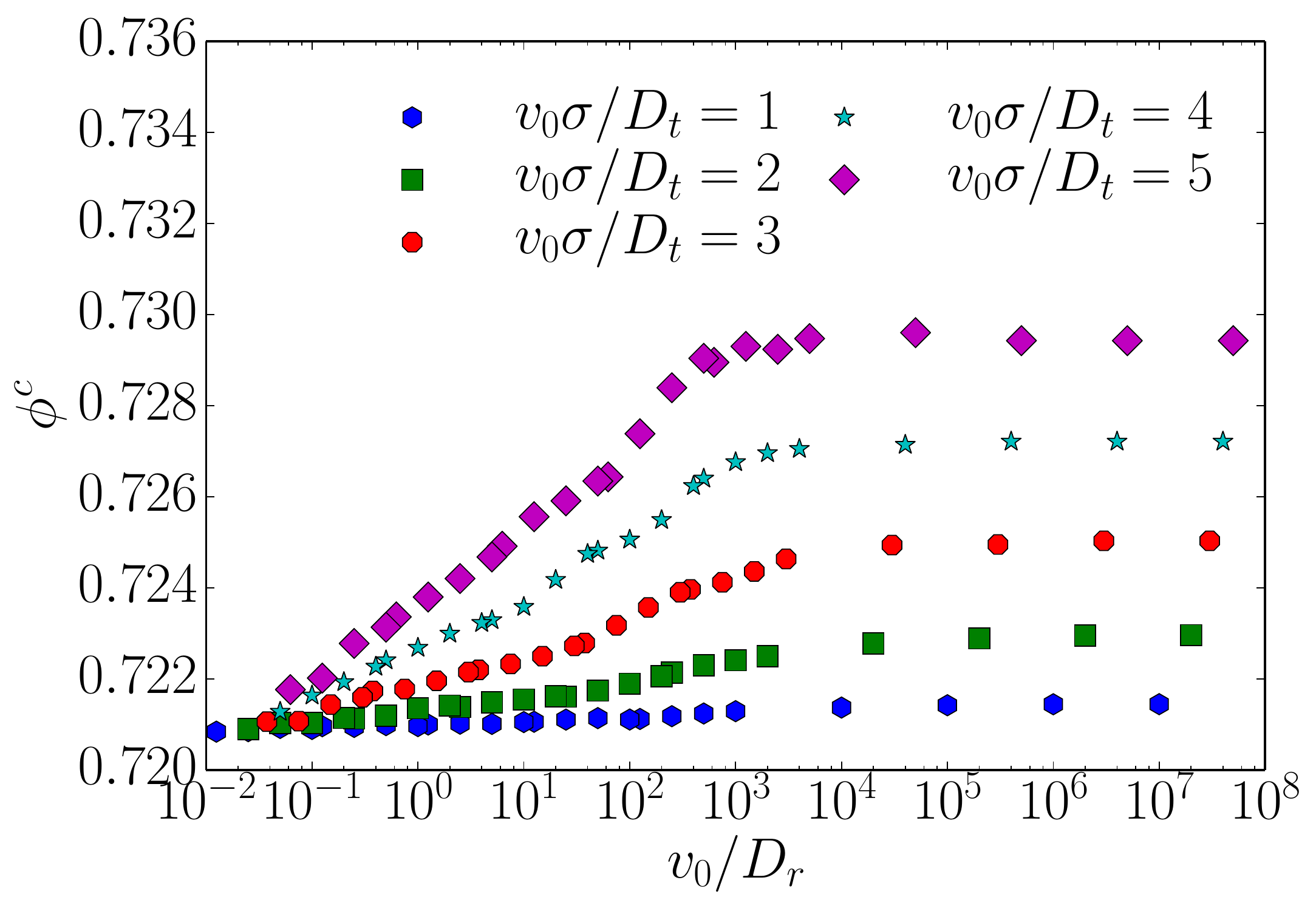}
\caption{\label{fig:phic_vs_dr}
  Critical packing fraction $\phi^{c}(v_{0},D_{r})$ as a function of
  persistence length $\ell_p=v_0/D_r$ for various self-propulsion speeds
  $v_0$ as labeled.
}
\end{figure}

Figure~\ref{fig:phic_vs_dr} demonstrates the role of the persistence
length on the glass transition. Here, the transition points $\phi^c$
obtained by power-law extrapolations are shown for various fixed $v_0$
as functions of $\ell_p$. In agreement with the discussion above,
the curves for different $v_0$ separate at large $\ell_p$. At fixed
persistence length, stronger self propulsion is more effective in
shifting the glass transition to higher densities. As $\ell_p$ approaches
zero, the glass-transition point of the passive system is recovered
for all $v_0$. Interestingly, the point where the activity-dependence
of the glass transition starts being significant is close to the point
where $\ell_p\approx\ell_c\approx0.1\sigma$, i.e., the cage size.

\begin{figure}
\includegraphics[width=\linewidth]{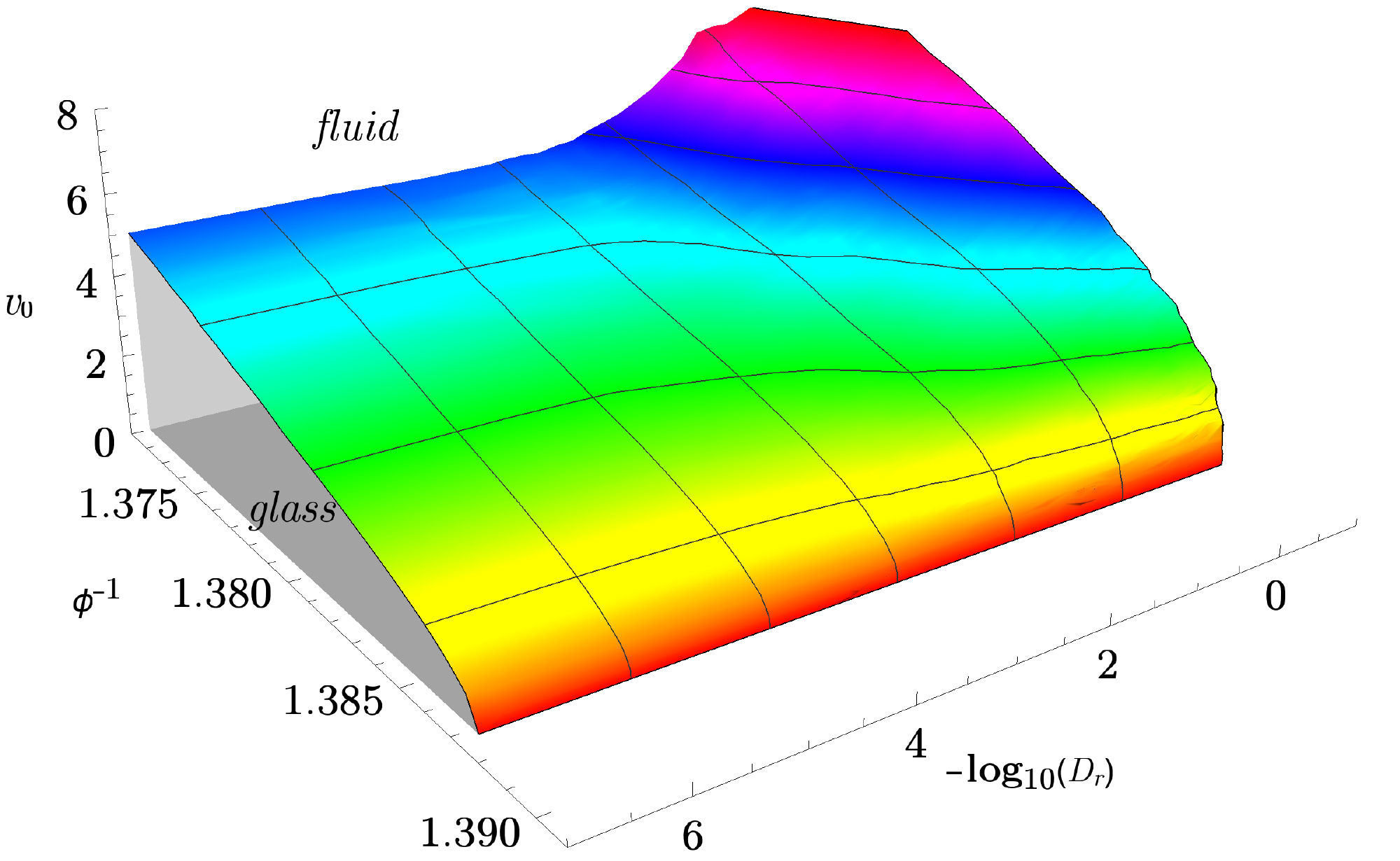}
\caption{\label{fig:3dplot}
  MCT glass-transition surface for hard-disk ABP, as estimated from
  power-law fits of the structural relaxation time. The grey shaded
  area below the surface is the glassy region.
}
\end{figure}

Figure~\ref{fig:3dplot} summarizes the estimated glass-transition points
of the MCT transition of active hard disks as a critical surface in
the parameter space spanned by $(1/\phi,\log_{10}(1/D_r),v_0)$.
This choice is motivated by a recent study of self-propelled cells in
a self-propelled Voronoi fluid model by Bi et~al.\ \cite{biophys:Bi.2016}.
In this model, a
self-adhesion parameter $p$ plays the role of an inverse density.
In Ref.~\cite{biophys:Bi.2016}, an active-glass diagram was conjectured that includes
two limiting shapes for $D_r\to0$ and $D_r\to\infty$ with a cross-over
between them around $D_r=1$. The numerical results shown in
Fig.~\ref{fig:3dplot} are in good agreement with the conjecture by
Bi et~al.\ \cite{biophys:Bi.2016}
over the range of $D_r$ shown. A glass-transition surface emerges
that extends from a limiting line in the $(v_0,1/\phi)$ plane at
$D_r=0$ and bends upwards to higher $v_0$ as $D_r$ increases beyond
$D_r\approx0.01$.
From the discussion above, one expects the
glass-transition surface to bend over to a vertical plane as
$D_r\to\infty$ (i.e., to the right of Fig.~\ref{fig:3dplot}),
different from what was suggested in Ref.~\cite{biophys:Bi.2016}.

Different from the case $D_r\to\infty$, the present MCT predicts
a $v_0$-dependent limiting shape of the glass transition in the limit
$D_r\to0$. This glass transition is given by the bifurcation points
of Eq.~\eqref{eq:mf0}.

\begin{figure}
\includegraphics[width=\linewidth]{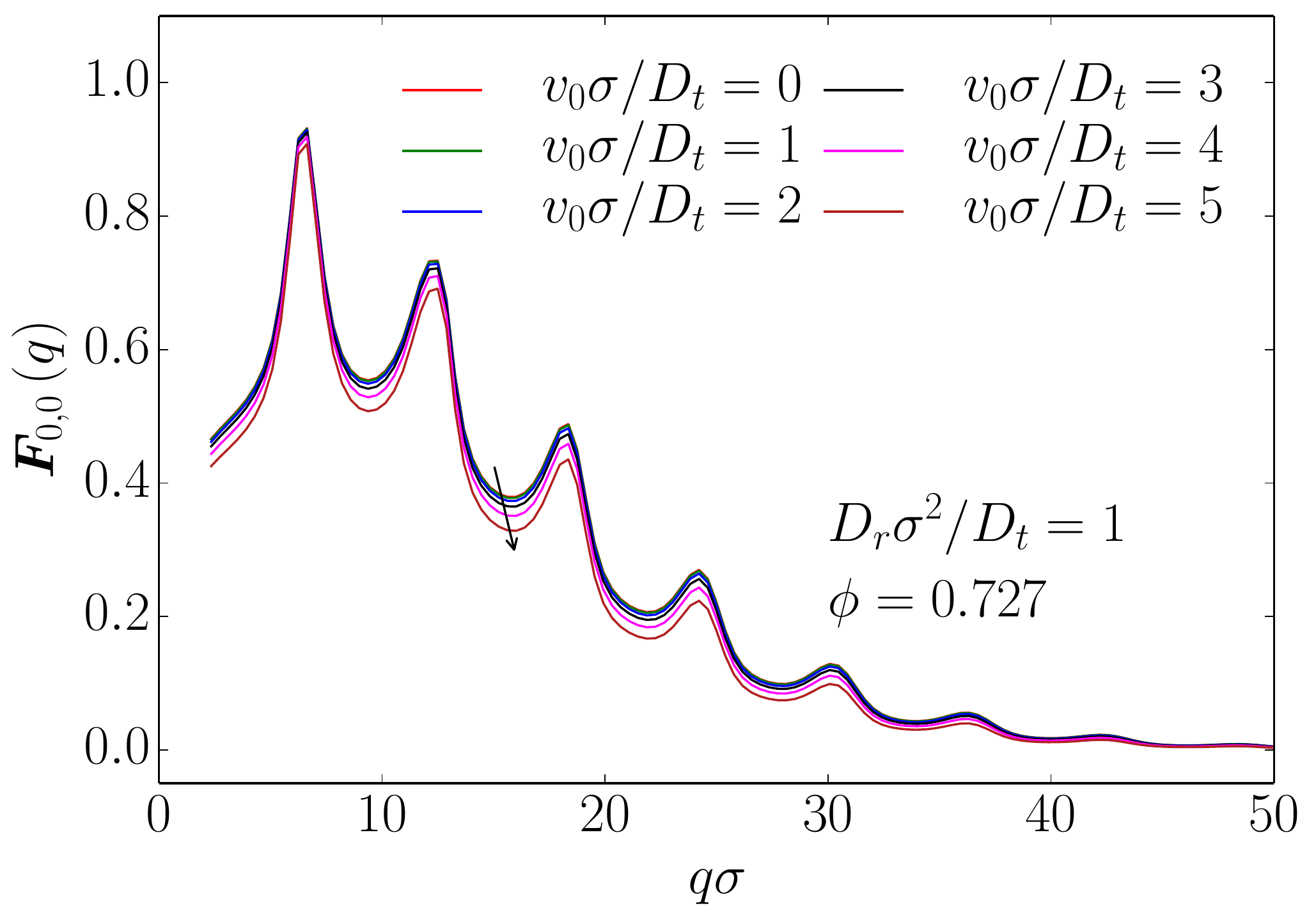}
\caption{\label{fig:fq}
  Nonergodicity parameters $F_{00}(\vec q)=\lim_{t\to\infty}\Phi_{00}(\vec q,t)$
  of the transient density correlation functions \TODO{how obtained}.
  Curves from top to bottom correspond to increasing self-propulsion
  velocity (as labeled) at fixed packing fraction $\phi=0.727$ and
  $D_r=1\,D_t/\sigma^2$.
}
\end{figure}

To characterize the glassy structure, we show in
Fig.~\ref{fig:fq} the nonergodicity parameters of the positional-density
correlation functions, $F_{00}(\vec q)$, for different $v_0$ at fixed
packing fraction. The passive case $v_0=0$ is included for reference;
it shows the known features of $F_{00}(\vec q)$: the nonergodicity
paraeters oscillate in phase with the static structure factor. They are
most pronounced around $q\sigma\approx7$, indicating that the glass is
stiffest with respect to density fluctuations of wavelengths comparable to the
particle size.

Increasing $v_0$, the nonergodicity parameters decrease for all $q$.
Thus, active driving renders the glass mechanically softer at fixed
density. The effect is however minor: for most $q$, the values for
$v_0=5\,D_t/\sigma$ are less than $10\%$ smaller compared to the passive
case.
The decrease in mechanical stiffness of the glass with increasing
activity might be counter-intuitive: with increasing P\'eclet number
$\tlname{Pe}$, the effective pressure of the system according to its
low-density description increases \cite{biophys:Tailleur.2008}. \TODO{check citations}
Such an increase in pressure might, by analogy to the passive system,
be expected to cause an increase in mechanical stiffness.
This shows that active forces act quite differently from thermodynamic
ones at the glass transition.

\begin{figure}
\includegraphics[width=\linewidth]{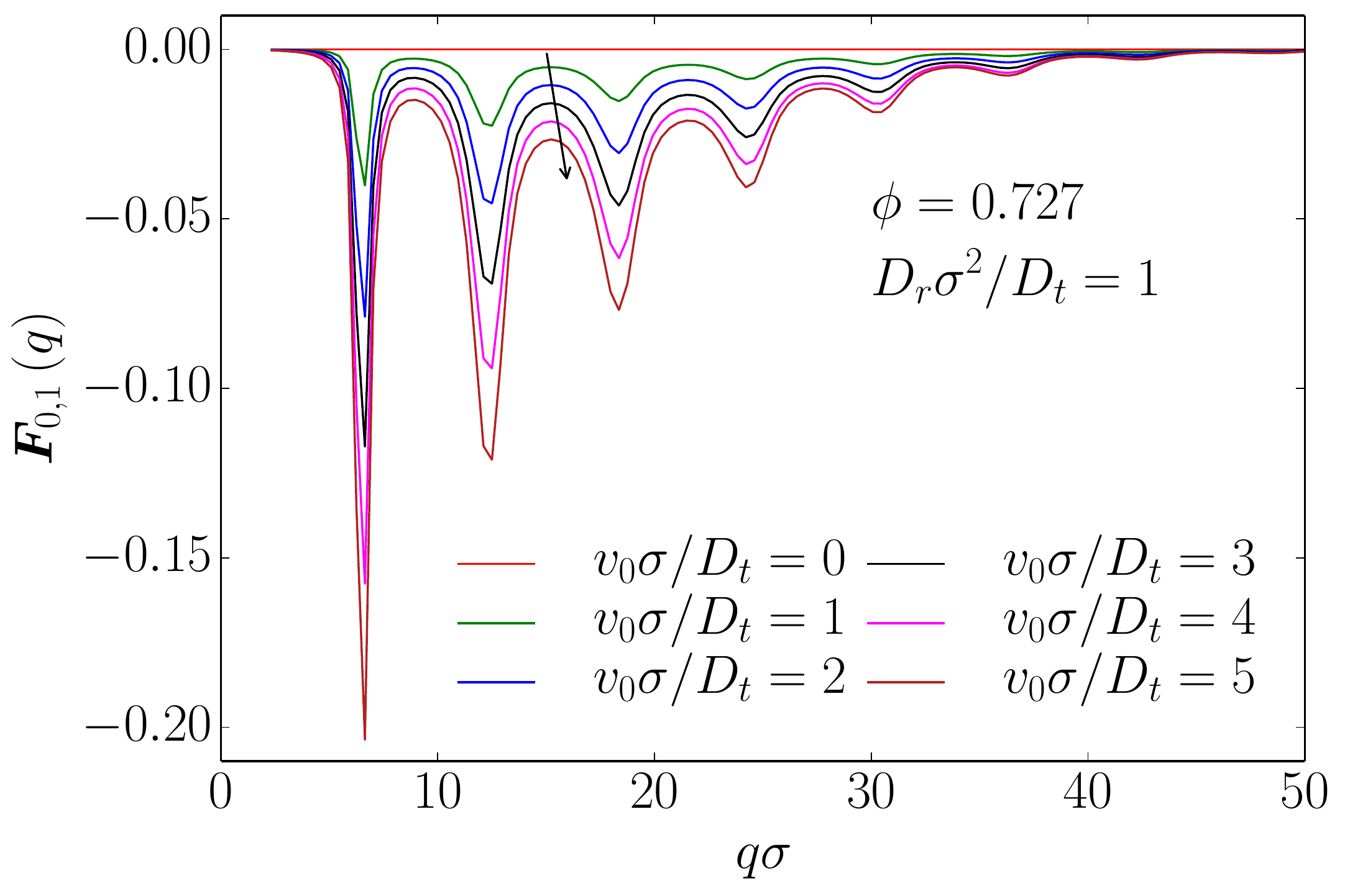}
\caption{\label{fig:fq1}
  Nonergodicity parameters $F_{01}(\vec q)$ of the orientation-translation
  coupling, for the same parameters as used in Fig.~\ref{fig:fq}.
}
\end{figure}

There is an interesting observation regarding the off-diagonal components
of $\bs F(\vec q)$: while $F_{l0}(\vec q)=0$ for all $l\neq0$, due to
the presence of the hopping term $\propto l^2D_r$, the MCT equations admit
solutions where $F_{0l}(\vec q)\neq0$ for all $l$. In particular
$F_{01}(\vec q)\neq0$, as shown in Fig.~\ref{fig:fq}.
In other words, the active glass keeps infinite memory not only in
the translational degrees of freedom, but also in the coupling of
orientations to translations. Loosely speaking, an initial orientation
fluctuation leaves its fingerprint in the positions in the glass,
while of course an initial density fluctuation does not affect the
later orientation fluctuations.

The non-trivial long-time limits $F_{01}(\vec q)$
are shown in Fig.~\ref{fig:fq1} for
the same state points as those in Fig.~\ref{fig:fq}.
For the reference frame chosen in our discussion, the values of
$F_{01}(\vec q)$ are negative and real; note that this is not invariant
under rotations of the coordinate system.
Apart from this, the $F_{01}(\vec q)$ show behavior that is qualitatively
similar to the one seen in $F_{00}(\vec q)$: oscillations are dictated by
those in the equilibrium static structure factor, and the strongest
conribution comes from nearest-neighbor cage distances.

\begin{figure}
\includegraphics[width=\linewidth]{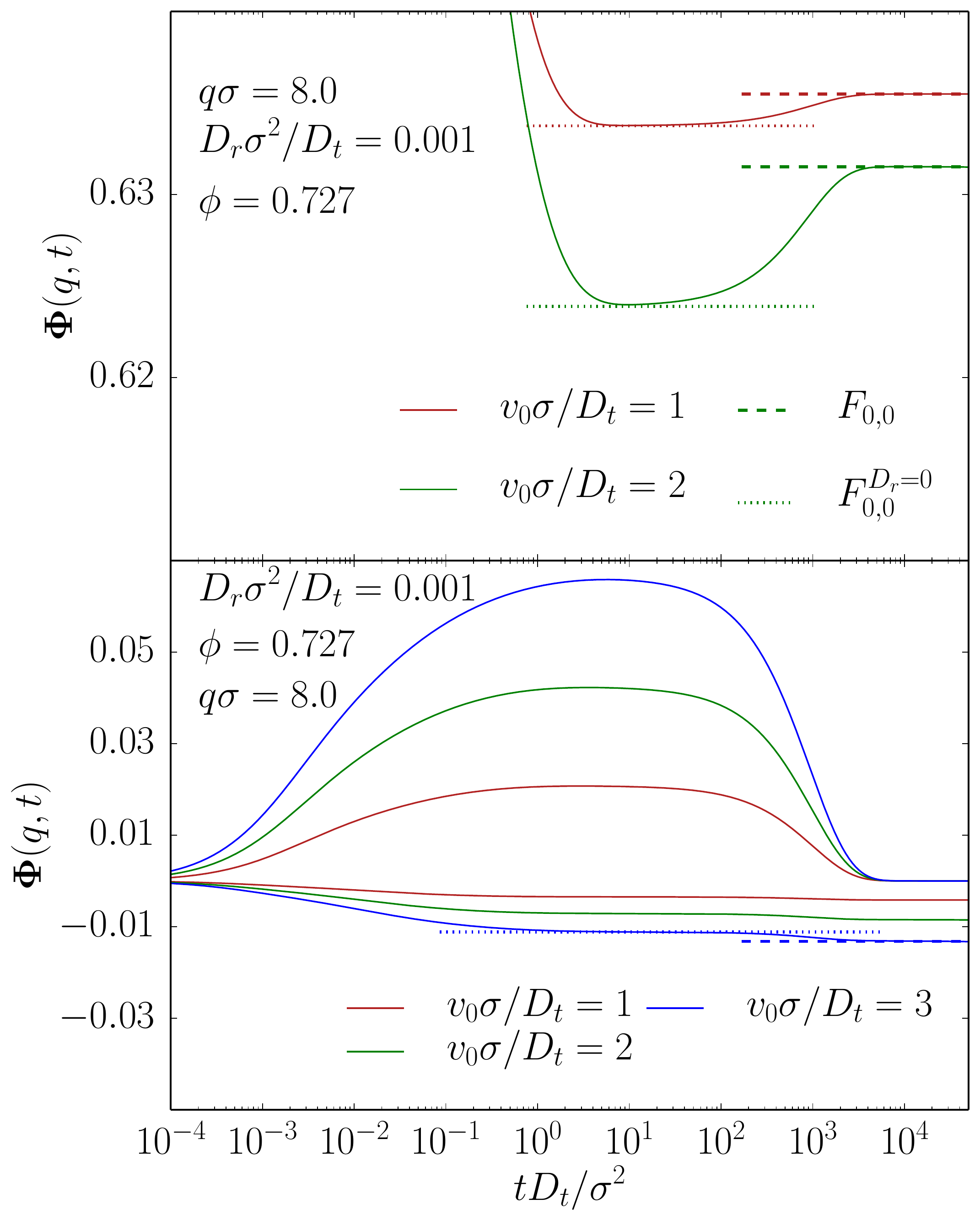}
\caption{\label{fig:two_plateaus}
Density correlator $\bs\Phi_{0,0}$ (solid lines) as a function of time for packing fraction above $\phi^{c}$ and different
velocities as labeled, for the case $D_r=1/1000\,D_t/\sigma^2$.
The dashed and dotted lines correspond to the nonergodicity parameters at finite and zero rotational diffusion, Eqs.~\eqref{eq:mf1} and \eqref{eq:mf0}, respectively.
}
\end{figure}

The nonergodicity parameters discussed above are the values attained by
the transient density correlation functions for
large times, i.e., $D_tt/\sigma^2\gg1$ and in particular $D_rt\gg1$.
In the case of strongly persistent motion, $D_r\ll D_t/\sigma^2$,
an intermediate time window opens where $\sigma^2/D_t\ll t\ll 1/D_r$,
i.e., for times large compared to the free diffusion time, but small
compared to the timescale of reorientations.

The existence of an additional slow time scale $\tau_\text{rot}=1/D_r$
has a pronounced effect on the density correlation functions.
The effect is highlighted by Fig.~\ref{fig:two_plateaus}:
a time window appears for times that are
large compared to the intrinsic translational time scale,
but still small compared to the reorientation time scale.
The density correlation functions approach a plateau
in this regime, that is given by the $D_r=0$ glass transition
Eq.~\eqref{eq:mf0}.
At $D_rt=1$, an exponential crossover is seen from this plateau
to a plateau given by the augmented equation for $D_r\neq0$,
Eq.~\eqref{eq:mf1}. This plateau depends on
$D_r$, even if the MCT memory kernel does not, because in
Eq.~\eqref{eq:mf1} there appears an integral over the full correlator.
Note that this radically changes the scaling laws known from standard
MCT: in the passive case, the final plateau in the ideal glass is
approached by a slow power law, the MCT critical law $t^{-a}$ (with
a nontrivial exponent $0<a<1/2$). In the present case, this critical
decay is only seen in the approach to the first plateau.
Hence the small-$D_r$ case displays the cross-over from the standard
MCT glass to an active glass governed by details of the short-time motion.
The ABP with small rotational diffusivity could thus serve as an
interesting model system to study the differences between two kinds
of glasses.

\section{Conclusions}\label{sec:conclusions}

We have developed a mode-coupling theory of the glass transition (MCT) of
active Brownian particles (ABP), for the special case of spherically symmetric
steric interactions. In contrast to prior approaches, we treat both the
translational and reorientational degrees of freedom of the ABP model
explicitly in the theory. This allows to cover the full parameter range
of self-propulsion velocities $v_0$ and persistence lengths $\ell_p=v_0/D_r$.
While at low densities, both parameters only enter through their combination,
$\tlname{Pe}=v_0\ell_p/D_t$, the high-density dynamics depends on both
parameters separately, because transient caging of particles imposes the
average cage size as an additional length scale on the system that interferes
with the persistence length.

The effects of active driving on the glassy dynamics are intuitive:
self propulsion enhances structural relaxation, but only if it is persistent
enough. Above a critical packing fraction $\phi_c(v_0,D_r)$, structural
relaxation by the MCT mechanism becomes ineffective, and for densities
above $\phi_c$, a glassy state is reached. The features of this glass
transition are in agreement with earlier simulation studies of the
three-dimensional analog of the hard-disk ABP model \cite{biophys:Ni.2013}.
Qualitatively, the predicted ABP-MCT glass transition also matches the
one that was estimated from earlier simulations of a self-propelled Voronoi
fluid model \cite{biophys:Bi.2016}. It is interesting to see that the qualitative analogies
between this foam-like model of slow dynamics and the hard-sphere model
extend from the passive case \TODO{cite strasbourg} to the respective self-propelled
versions. In this sense, the ABP model is closer to the Voronoi fluid model
than to the athermal AOUP model whose active glass has been discussed in
the context of MCT.

That the glass transition persists under finite activity is not a priori
clear. A different scenario arises, for example, for shear-driven passive
colloidal suspensions where MCT predicts the glass transition to vanish for
arbitrarily weak driving \cite{rheology:Fuchs.2002c}. This emphasizes that there are
qualitatively different ways of driving a dense system out of equilibrium.
Note however that according to our numerical results, the glass transition
surface is a smooth surface around $v_0=0$; in particular, any small
activity just at the passive glass transition will destroy the glass.
This holds even in the case $D_r=0$, i.e., if each particle maintains
its randomly chosen initial propulsion direction forever.
This is at odds with the application of MCT to active microrheology of
passive suspensions, i.e., the case of a single persistently driven particle
in a bath of passive ones \cite{rheology:Gazuz.2009}. There it was
found that a finite force threshold needs to be overcome to delocalize
the driven particle. This threshold was associated to the strength of
nearest-neighbor cages. It appears that collective driving of all particles
allows to collectively break such cages, so that the force required to do
so by a single particle approaches zero.

The ABP-MCT predicts a surface of idealized glass transitions in the
parameter space spanned by the packing fraction, the persistence length,
and the self-propulsion velocity $(\phi,\ell_p,v_0)$. This surface does
not collapse onto a single line in the $(\phi,\tlname{Pe})$ plane as predicted
by theories that start from a coarse-grained description of the orientational
degrees of freedom. Since the P\'eclet number $\tlname{Pe}$ enters the
definition of an effective temperature that can be used to describe many
properties of the dilute system
\TODO{check citations, add Brady} \cite{biophys:Palacci.2010,biophys:Enculescu.2011,biophys:Wang.2011},
this may be taken as an indication that the mapping of the glassy dynamics
of ABP onto that of a near-equilibrium system should proceed with care.
In this sense, the situation is similar to that encountered close
to phase transitions driven by active motion
\TODO{check citations}
\cite{biophys:Tailleur.2008,biophys:Tailleur.2009,biophys:Bialke.2012}.

Our results emphasize the interplay between the structural caging length
scale typical for glass-forming systems, and the persistence length associated
to active swimming. If the persistence length is smaller than the caging
length, activity does not shift the glass transition.

In the glass, density-density correlation functions ($S_{00}(\vec q,t)$)
do not decay to zero, but to a finite long-time limit. This finite
nonergodicity parameter quantifies the overlap of an initial positional
density fluctuation after infinitely long propagation with the positional
density fluctuations themselves. Similarly, a non-zero long-time limit
emerges for $S_{0l}(\vec q,t)$: also the ininitely-long propagated
orientational fluctuations are required to determine the statistics of
positional fluctuations in the glass. In this sense, the active glass
keeps a memory of both initial positions and initial orientations.

The passive glass according to MCT is characterized by quantities
that do not depend on the parameters determining the short-time motion;
for a passive Brownian hard-sphere system, the short-time diffusivities
$D_t$ and $D_r$ are irrelevant in the glass. This is no longer true
for the active glass: here, the equation that determines the MCT
glass transition depends on the integral of the correlation function,
and thus on the details of short-time diffusion in principle. In this
sense, the active glass is quite different from the passive one.

Our ABP-MCT is based on the integration-through transients
(ITT) formalism, and thus focuses on the calculation of so-called transient
correlation functions that are formed with the full non-equilibrium dynamics
but averaged using the equilibrium Boltzmann weight. These correlation
functions are natural starting points for the calculation of non-equilibrium
averages of, in principle, arbitrary observables within ITT. This will
be the focus in future work.

\TODO{conclusions: there are two relevant P\'eclet numbers,
$\tlname{Pe}_a=v_0\sigma/D_\text{act}=\sigma D_r/v_0=\sigma/\ell_p$ the
activity P\'eclet number and the translational P\'eclet number
$\tlname{Pe}_t=v_0\sigma/D_t$. Note that $\tlname{Pe}_r=v_0/\sigma D_r
=1/\tlname{Pe}_a$ could also be used as a rotational P\'eclet number.}

\begin{acknowledgments}
We acknowledge funding from Deutsche Forschungsgemeinschaft (DFG), as part of the Special Priority Programme SPP~1726 ``Microswimmers''.
The authors gratefully acknowledge the computing time granted by the John von Neumann Institute for Computing (NIC) and provided on the supercomputer JURECA \cite{jureca} at Jülich Supercomputing Centre (JSC).
\TODO{Need to send them a copy of the final paper.}
We thank T.~Franosch and R.~Schilling for helpful discussions.
\end{acknowledgments}

\begin{appendix}

\section{Derivation of MCT Vertex}\label{sec:mctderiv}

\section{Transformed MCT Equations}\label{sec:mctnum}

\section{Numerical Algorithm}\label{sec:mctalgo}

\end{appendix}

\bibliography{biophys,rheology,glass-theory,bibtex-jureca-txt-file,add}

\begin{thebibliography}{55}%
\makeatletter
\providecommand \@ifxundefined [1]{%
 \@ifx{#1\undefined}
}%
\providecommand \@ifnum [1]{%
 \ifnum #1\expandafter \@firstoftwo
 \else \expandafter \@secondoftwo
 \fi
}%
\providecommand \@ifx [1]{%
 \ifx #1\expandafter \@firstoftwo
 \else \expandafter \@secondoftwo
 \fi
}%
\providecommand \natexlab [1]{#1}%
\providecommand \enquote  [1]{``#1''}%
\providecommand \bibnamefont  [1]{#1}%
\providecommand \bibfnamefont [1]{#1}%
\providecommand \citenamefont [1]{#1}%
\providecommand \href@noop [0]{\@secondoftwo}%
\providecommand \href [0]{\begingroup \@sanitize@url \@href}%
\providecommand \@href[1]{\@@startlink{#1}\@@href}%
\providecommand \@@href[1]{\endgroup#1\@@endlink}%
\providecommand \@sanitize@url [0]{\catcode `\\12\catcode `\$12\catcode
  `\&12\catcode `\#12\catcode `\^12\catcode `\_12\catcode `\%12\relax}%
\providecommand \@@startlink[1]{}%
\providecommand \@@endlink[0]{}%
\providecommand \url  [0]{\begingroup\@sanitize@url \@url }%
\providecommand \@url [1]{\endgroup\@href {#1}{\urlprefix }}%
\providecommand \urlprefix  [0]{URL }%
\providecommand \Eprint [0]{\href }%
\providecommand \doibase [0]{http://dx.doi.org/}%
\providecommand \selectlanguage [0]{\@gobble}%
\providecommand \bibinfo  [0]{\@secondoftwo}%
\providecommand \bibfield  [0]{\@secondoftwo}%
\providecommand \translation [1]{[#1]}%
\providecommand \BibitemOpen [0]{}%
\providecommand \bibitemStop [0]{}%
\providecommand \bibitemNoStop [0]{.\EOS\space}%
\providecommand \EOS [0]{\spacefactor3000\relax}%
\providecommand \BibitemShut  [1]{\csname bibitem#1\endcsname}%
\let\auto@bib@innerbib\@empty
\bibitem [{\citenamefont {{\relax Poujade}}\ \emph {et~al.}(2007)\citenamefont
  {{\relax Poujade}}, \citenamefont {{\relax Grasland-Mongrain}}, \citenamefont
  {{\relax Hertzog}}, \citenamefont {{\relax Jouanneau}}, \citenamefont
  {{\relax Chavrier}}, \citenamefont {{\relax Ladoux}}, \citenamefont {{\relax
  Buguin}},\ and\ \citenamefont {{\relax Silberzan}}}]{biophys:Poujade.2007}%
  \BibitemOpen
  \bibfield  {author} {\bibinfo {author} {\bibfnamefont {M.}~\bibnamefont
  {{\relax Poujade}}}, \bibinfo {author} {\bibfnamefont {E.}~\bibnamefont
  {{\relax Grasland-Mongrain}}}, \bibinfo {author} {\bibfnamefont
  {A.}~\bibnamefont {{\relax Hertzog}}}, \bibinfo {author} {\bibfnamefont
  {J.}~\bibnamefont {{\relax Jouanneau}}}, \bibinfo {author} {\bibfnamefont
  {P.}~\bibnamefont {{\relax Chavrier}}}, \bibinfo {author} {\bibfnamefont
  {B.}~\bibnamefont {{\relax Ladoux}}}, \bibinfo {author} {\bibfnamefont
  {A.}~\bibnamefont {{\relax Buguin}}}, \ and\ \bibinfo {author} {\bibfnamefont
  {P.}~\bibnamefont {{\relax Silberzan}}},\ }\href@noop {} {\bibfield
  {journal} {\bibinfo  {journal} {Proc. Natl. Acad. Sci. USA}\ }\textbf
  {\bibinfo {volume} {104}},\ \bibinfo {pages} {15988} (\bibinfo {year}
  {2007})}\BibitemShut {NoStop}%
\bibitem [{\citenamefont {{\relax Petitjean}}\ \emph
  {et~al.}(2010)\citenamefont {{\relax Petitjean}}, \citenamefont {{\relax
  Reffay}}, \citenamefont {{\relax Grasland-Mongrain}}, \citenamefont {{\relax
  Poujade}}, \citenamefont {{\relax Ladoux}}, \citenamefont {{\relax Buguin}},\
  and\ \citenamefont {{\relax Silberzan}}}]{biophys:Petitjean.2010}%
  \BibitemOpen
  \bibfield  {author} {\bibinfo {author} {\bibfnamefont {L.}~\bibnamefont
  {{\relax Petitjean}}}, \bibinfo {author} {\bibfnamefont {M.}~\bibnamefont
  {{\relax Reffay}}}, \bibinfo {author} {\bibfnamefont {E.}~\bibnamefont
  {{\relax Grasland-Mongrain}}}, \bibinfo {author} {\bibfnamefont
  {M.}~\bibnamefont {{\relax Poujade}}}, \bibinfo {author} {\bibfnamefont
  {B.}~\bibnamefont {{\relax Ladoux}}}, \bibinfo {author} {\bibfnamefont
  {A.}~\bibnamefont {{\relax Buguin}}}, \ and\ \bibinfo {author} {\bibfnamefont
  {P.}~\bibnamefont {{\relax Silberzan}}},\ }\href@noop {} {\bibfield
  {journal} {\bibinfo  {journal} {Biophys. J.}\ }\textbf {\bibinfo {volume}
  {98}},\ \bibinfo {pages} {1790} (\bibinfo {year} {2010})}\BibitemShut
  {NoStop}%
\bibitem [{\citenamefont {{\relax Trepat}}\ \emph {et~al.}(2009)\citenamefont
  {{\relax Trepat}}, \citenamefont {{\relax Wasserman}}, \citenamefont {{\relax
  Angelini}}, \citenamefont {{\relax Millet}}, \citenamefont {{\relax Weitz}},
  \citenamefont {{\relax Butler}},\ and\ \citenamefont {{\relax
  Fredberg}}}]{biophys:Trepat.2009}%
  \BibitemOpen
  \bibfield  {author} {\bibinfo {author} {\bibfnamefont {X.}~\bibnamefont
  {{\relax Trepat}}}, \bibinfo {author} {\bibfnamefont {M.~R.}\ \bibnamefont
  {{\relax Wasserman}}}, \bibinfo {author} {\bibfnamefont {T.~E.}\ \bibnamefont
  {{\relax Angelini}}}, \bibinfo {author} {\bibfnamefont {E.}~\bibnamefont
  {{\relax Millet}}}, \bibinfo {author} {\bibfnamefont {D.~A.}\ \bibnamefont
  {{\relax Weitz}}}, \bibinfo {author} {\bibfnamefont {J.~P.}\ \bibnamefont
  {{\relax Butler}}}, \ and\ \bibinfo {author} {\bibfnamefont {J.~J.}\
  \bibnamefont {{\relax Fredberg}}},\ }\href@noop {} {\bibfield  {journal}
  {\bibinfo  {journal} {Nature Phys.}\ }\textbf {\bibinfo {volume} {5}},\
  \bibinfo {pages} {426} (\bibinfo {year} {2009})}\BibitemShut {NoStop}%
\bibitem [{\citenamefont {{\relax Angelini}}\ \emph {et~al.}(2010)\citenamefont
  {{\relax Angelini}}, \citenamefont {{\relax Hannezo}}, \citenamefont {{\relax
  Trepat}}, \citenamefont {{\relax Fredberg}},\ and\ \citenamefont {{\relax
  Weitz}}}]{biophys:Angelini.2010}%
  \BibitemOpen
  \bibfield  {author} {\bibinfo {author} {\bibfnamefont {T.~E.}\ \bibnamefont
  {{\relax Angelini}}}, \bibinfo {author} {\bibfnamefont {E.}~\bibnamefont
  {{\relax Hannezo}}}, \bibinfo {author} {\bibfnamefont {X.}~\bibnamefont
  {{\relax Trepat}}}, \bibinfo {author} {\bibfnamefont {J.~J.}\ \bibnamefont
  {{\relax Fredberg}}}, \ and\ \bibinfo {author} {\bibfnamefont {D.~A.}\
  \bibnamefont {{\relax Weitz}}},\ }\href@noop {} {\bibfield  {journal}
  {\bibinfo  {journal} {Phys. Rev. Lett.}\ }\textbf {\bibinfo {volume}
  {104}},\ \bibinfo {pages} {168104} (\bibinfo {year} {2010})}\BibitemShut
  {NoStop}%
\bibitem [{\citenamefont {{\relax Bi}}\ \emph {et~al.}(2016)\citenamefont
  {{\relax Bi}}, \citenamefont {{\relax Yang}}, \citenamefont {{\relax
  Marchetti}},\ and\ \citenamefont {{\relax Manning}}}]{biophys:Bi.2016}%
  \BibitemOpen
  \bibfield  {author} {\bibinfo {author} {\bibfnamefont {D.}~\bibnamefont
  {{\relax Bi}}}, \bibinfo {author} {\bibfnamefont {X.}~\bibnamefont {{\relax
  Yang}}}, \bibinfo {author} {\bibfnamefont {M.~C.}\ \bibnamefont {{\relax
  Marchetti}}}, \ and\ \bibinfo {author} {\bibfnamefont {M.~L.}\ \bibnamefont
  {{\relax Manning}}},\ }\href@noop {} {\bibfield  {journal} {\bibinfo
  {journal} {Phys. Rev. X}\ }\textbf {\bibinfo {volume} {6}},\ \bibinfo
  {pages} {021011} (\bibinfo {year} {2016})}\BibitemShut {NoStop}%
\bibitem [{\citenamefont {{\relax Peruani}}\ \emph {et~al.}(2012)\citenamefont
  {{\relax Peruani}}, \citenamefont {{\relax Starruß}}, \citenamefont {{\relax
  Jakovljevic}}, \citenamefont {{\relax Søgaard-Andersen}}, \citenamefont
  {{\relax Deutsch}},\ and\ \citenamefont {{\relax
  Bär}}}]{biophys:Peruani.2012}%
  \BibitemOpen
  \bibfield  {author} {\bibinfo {author} {\bibfnamefont {F.}~\bibnamefont
  {{\relax Peruani}}}, \bibinfo {author} {\bibfnamefont {J.}~\bibnamefont
  {{\relax Starruß}}}, \bibinfo {author} {\bibfnamefont {V.}~\bibnamefont
  {{\relax Jakovljevic}}}, \bibinfo {author} {\bibfnamefont {L.}~\bibnamefont
  {{\relax Søgaard-Andersen}}}, \bibinfo {author} {\bibfnamefont
  {A.}~\bibnamefont {{\relax Deutsch}}}, \ and\ \bibinfo {author}
  {\bibfnamefont {M.}~\bibnamefont {{\relax Bär}}},\ }\href@noop {} {\bibfield
   {journal} {\bibinfo  {journal} {Phys. Rev. Lett.}\ }\textbf {\bibinfo
  {volume} {108}},\ \bibinfo {pages} {098102} (\bibinfo {year}
  {2012})}\BibitemShut {NoStop}%
\bibitem [{\citenamefont {{\relax Fabry}}\ \emph {et~al.}(2001)\citenamefont
  {{\relax Fabry}}, \citenamefont {{\relax Maksym}}, \citenamefont {{\relax
  Butler}}, \citenamefont {{\relax Glogauer}}, \citenamefont {{\relax
  Navajas}},\ and\ \citenamefont {{\relax Fredberg}}}]{biophys:Fabry.2001}%
  \BibitemOpen
  \bibfield  {author} {\bibinfo {author} {\bibfnamefont {B.}~\bibnamefont
  {{\relax Fabry}}}, \bibinfo {author} {\bibfnamefont {G.~N.}\ \bibnamefont
  {{\relax Maksym}}}, \bibinfo {author} {\bibfnamefont {J.~P.}\ \bibnamefont
  {{\relax Butler}}}, \bibinfo {author} {\bibfnamefont {M.}~\bibnamefont
  {{\relax Glogauer}}}, \bibinfo {author} {\bibfnamefont {D.}~\bibnamefont
  {{\relax Navajas}}}, \ and\ \bibinfo {author} {\bibfnamefont {J.~J.}\
  \bibnamefont {{\relax Fredberg}}},\ }\href@noop {} {\bibfield  {journal}
  {\bibinfo  {journal} {Phys. Rev. Lett.}\ }\textbf {\bibinfo {volume}
  {87}},\ \bibinfo {pages} {148102} (\bibinfo {year} {2001})}\BibitemShut
  {NoStop}%
\bibitem [{\citenamefont {{\relax Bursac}}\ \emph {et~al.}(2005)\citenamefont
  {{\relax Bursac}}, \citenamefont {{\relax Lenormand}}, \citenamefont {{\relax
  Fabry}}, \citenamefont {{\relax Oliver}}, \citenamefont {{\relax Weitz}},
  \citenamefont {{\relax Viasnoff}}, \citenamefont {{\relax Butler}},\ and\
  \citenamefont {{\relax Fredberg}}}]{biophys:Bursac.2005}%
  \BibitemOpen
  \bibfield  {author} {\bibinfo {author} {\bibfnamefont {P.}~\bibnamefont
  {{\relax Bursac}}}, \bibinfo {author} {\bibfnamefont {G.}~\bibnamefont
  {{\relax Lenormand}}}, \bibinfo {author} {\bibfnamefont {B.}~\bibnamefont
  {{\relax Fabry}}}, \bibinfo {author} {\bibfnamefont {M.}~\bibnamefont
  {{\relax Oliver}}}, \bibinfo {author} {\bibfnamefont {D.~A.}\ \bibnamefont
  {{\relax Weitz}}}, \bibinfo {author} {\bibfnamefont {V.}~\bibnamefont
  {{\relax Viasnoff}}}, \bibinfo {author} {\bibfnamefont {J.~P.}\ \bibnamefont
  {{\relax Butler}}}, \ and\ \bibinfo {author} {\bibfnamefont {J.~J.}\
  \bibnamefont {{\relax Fredberg}}},\ }\href@noop {} {\bibfield  {journal}
  {\bibinfo  {journal} {Nature Materials}\ }\textbf {\bibinfo {volume} {4}},\
  \bibinfo {pages} {557} (\bibinfo {year} {2005})}\BibitemShut {NoStop}%
\bibitem [{\citenamefont {{\relax Wang}}\ \emph {et~al.}(2011)\citenamefont
  {{\relax Wang}}, \citenamefont {{\relax Shen}},\ and\ \citenamefont {{\relax
  Wolynes}}}]{biophys:Wang.2011b}%
  \BibitemOpen
  \bibfield  {author} {\bibinfo {author} {\bibfnamefont {S.}~\bibnamefont
  {{\relax Wang}}}, \bibinfo {author} {\bibfnamefont {T.}~\bibnamefont {{\relax
  Shen}}}, \ and\ \bibinfo {author} {\bibfnamefont {P.~G.}\ \bibnamefont
  {{\relax Wolynes}}},\ }\href@noop {} {\bibfield  {journal} {\bibinfo
  {journal} {J. Chem. Phys.}\ }\textbf {\bibinfo {volume} {134}},\ \bibinfo
  {pages} {014510} (\bibinfo {year} {2011})}\BibitemShut {NoStop}%
\bibitem [{\citenamefont {{\relax Erbe}}\ \emph {et~al.}(2008)\citenamefont
  {{\relax Erbe}}, \citenamefont {{\relax Zientara}}, \citenamefont {{\relax
  Baraban}}, \citenamefont {{\relax Kreidler}},\ and\ \citenamefont {{\relax
  Leiderer}}}]{rheology:Erbe.2008}%
  \BibitemOpen
  \bibfield  {author} {\bibinfo {author} {\bibfnamefont {A.}~\bibnamefont
  {{\relax Erbe}}}, \bibinfo {author} {\bibfnamefont {M.}~\bibnamefont {{\relax
  Zientara}}}, \bibinfo {author} {\bibfnamefont {L.}~\bibnamefont {{\relax
  Baraban}}}, \bibinfo {author} {\bibfnamefont {C.}~\bibnamefont {{\relax
  Kreidler}}}, \ and\ \bibinfo {author} {\bibfnamefont {P.}~\bibnamefont
  {{\relax Leiderer}}},\ }\href@noop {} {\bibfield  {journal} {\bibinfo
  {journal} {J. Phys.: Condens. Matter}\ }\textbf {\bibinfo {volume} {20}}
  (\bibinfo {year} {2008})}\BibitemShut {NoStop}%
\bibitem [{\citenamefont {{\relax Baraban}}\ \emph {et~al.}(2012)\citenamefont
  {{\relax Baraban}}, \citenamefont {{\relax Tasinkevych}}, \citenamefont
  {{\relax Popescu}}, \citenamefont {{\relax Sanchez}}, \citenamefont {{\relax
  Dietrich}},\ and\ \citenamefont {{\relax Schmidt}}}]{biophys:Baraban.2012}%
  \BibitemOpen
  \bibfield  {author} {\bibinfo {author} {\bibfnamefont {L.}~\bibnamefont
  {{\relax Baraban}}}, \bibinfo {author} {\bibfnamefont {M.}~\bibnamefont
  {{\relax Tasinkevych}}}, \bibinfo {author} {\bibfnamefont {M.~N.}\
  \bibnamefont {{\relax Popescu}}}, \bibinfo {author} {\bibfnamefont
  {S.}~\bibnamefont {{\relax Sanchez}}}, \bibinfo {author} {\bibfnamefont
  {S.}~\bibnamefont {{\relax Dietrich}}}, \ and\ \bibinfo {author}
  {\bibfnamefont {O.~G.}\ \bibnamefont {{\relax Schmidt}}},\ }\href@noop {}
  {\bibfield  {journal} {\bibinfo  {journal} {Soft Matter}\ }\textbf {\bibinfo
  {volume} {8}},\ \bibinfo {pages} {48} (\bibinfo {year} {2012})}\BibitemShut
  {NoStop}%
\bibitem [{\citenamefont {{\relax Palacci}}\ \emph {et~al.}(2010)\citenamefont
  {{\relax Palacci}}, \citenamefont {{\relax Cottin-Bizonne}}, \citenamefont
  {{\relax Ybert}},\ and\ \citenamefont {{\relax
  Bocquet}}}]{biophys:Palacci.2010}%
  \BibitemOpen
  \bibfield  {author} {\bibinfo {author} {\bibfnamefont {J.}~\bibnamefont
  {{\relax Palacci}}}, \bibinfo {author} {\bibfnamefont {C.}~\bibnamefont
  {{\relax Cottin-Bizonne}}}, \bibinfo {author} {\bibfnamefont
  {C.}~\bibnamefont {{\relax Ybert}}}, \ and\ \bibinfo {author} {\bibfnamefont
  {L.}~\bibnamefont {{\relax Bocquet}}},\ }\href@noop {} {\bibfield  {journal}
  {\bibinfo  {journal} {Phys. Rev. Lett.}\ }\textbf {\bibinfo {volume}
  {105}},\ \bibinfo {pages} {088304} (\bibinfo {year} {2010})}\BibitemShut
  {NoStop}%
\bibitem [{\citenamefont {{\relax Buttinoni}}\ \emph
  {et~al.}(2013)\citenamefont {{\relax Buttinoni}}, \citenamefont {{\relax
  Bialké}}, \citenamefont {{\relax Kümmel}}, \citenamefont {{\relax Löwen}},
  \citenamefont {{\relax Bechinger}},\ and\ \citenamefont {{\relax
  Speck}}}]{biophys:Buttinoni.2013}%
  \BibitemOpen
  \bibfield  {author} {\bibinfo {author} {\bibfnamefont {I.}~\bibnamefont
  {{\relax Buttinoni}}}, \bibinfo {author} {\bibfnamefont {J.}~\bibnamefont
  {{\relax Bialké}}}, \bibinfo {author} {\bibfnamefont {F.}~\bibnamefont
  {{\relax Kümmel}}}, \bibinfo {author} {\bibfnamefont {H.}~\bibnamefont
  {{\relax Löwen}}}, \bibinfo {author} {\bibfnamefont {C.}~\bibnamefont
  {{\relax Bechinger}}}, \ and\ \bibinfo {author} {\bibfnamefont
  {T.}~\bibnamefont {{\relax Speck}}},\ }\href@noop {} {\bibfield  {journal}
  {\bibinfo  {journal} {Phys. Rev. Lett.}\ }\textbf {\bibinfo {volume}
  {110}},\ \bibinfo {pages} {238301} (\bibinfo {year} {2013})}\BibitemShut
  {NoStop}%
\bibitem [{\citenamefont {{\relax Volpe}}\ \emph {et~al.}(2011)\citenamefont
  {{\relax Volpe}}, \citenamefont {{\relax Buttinoni}}, \citenamefont {{\relax
  Vogt}}, \citenamefont {{\relax Kümmerer}},\ and\ \citenamefont {{\relax
  Bechinger}}}]{biophys:Volpe.2011}%
  \BibitemOpen
  \bibfield  {author} {\bibinfo {author} {\bibfnamefont {G.}~\bibnamefont
  {{\relax Volpe}}}, \bibinfo {author} {\bibfnamefont {I.}~\bibnamefont
  {{\relax Buttinoni}}}, \bibinfo {author} {\bibfnamefont {D.}~\bibnamefont
  {{\relax Vogt}}}, \bibinfo {author} {\bibfnamefont {H.-J.}\ \bibnamefont
  {{\relax Kümmerer}}}, \ and\ \bibinfo {author} {\bibfnamefont
  {C.}~\bibnamefont {{\relax Bechinger}}},\ }\href@noop {} {\bibfield
  {journal} {\bibinfo  {journal} {Soft Matter}\ }\textbf {\bibinfo {volume}
  {7}},\ \bibinfo {pages} {8810} (\bibinfo {year} {2011})}\BibitemShut
  {NoStop}%
\bibitem [{\citenamefont {{\relax Zöttl}}\ and\ \citenamefont {{\relax
  Stark}}(2016)}]{biophys:Zoettl.2016}%
  \BibitemOpen
  \bibfield  {author} {\bibinfo {author} {\bibfnamefont {A.}~\bibnamefont
  {{\relax Zöttl}}}\ and\ \bibinfo {author} {\bibfnamefont {H.}~\bibnamefont
  {{\relax Stark}}},\ }\href@noop {} {\bibfield  {journal} {\bibinfo  {journal}
  {J. Phys.: Condens. Matter}\ }\textbf {\bibinfo {volume} {28}},\ \bibinfo
  {pages} {253001} (\bibinfo {year} {2016})}\BibitemShut {NoStop}%
\bibitem [{\citenamefont {{\relax Marchetti}}\ \emph
  {et~al.}(2016)\citenamefont {{\relax Marchetti}}, \citenamefont {{\relax
  Fily}}, \citenamefont {{\relax Henkes}}, \citenamefont {{\relax Patch}},\
  and\ \citenamefont {{\relax Yllanes}}}]{biophys:Marchetti.2016}%
  \BibitemOpen
  \bibfield  {author} {\bibinfo {author} {\bibfnamefont {M.~C.}\ \bibnamefont
  {{\relax Marchetti}}}, \bibinfo {author} {\bibfnamefont {Y.}~\bibnamefont
  {{\relax Fily}}}, \bibinfo {author} {\bibfnamefont {S.}~\bibnamefont {{\relax
  Henkes}}}, \bibinfo {author} {\bibfnamefont {A.}~\bibnamefont {{\relax
  Patch}}}, \ and\ \bibinfo {author} {\bibfnamefont {D.}~\bibnamefont {{\relax
  Yllanes}}},\ }\href@noop {} {\bibfield  {journal} {\bibinfo  {journal}
  {Curr. Opin. Colloid Interf. Sci.}\ }\textbf {\bibinfo {volume} {21}},\
  \bibinfo {pages} {34} (\bibinfo {year} {2016})}\BibitemShut {NoStop}%
\bibitem [{\citenamefont {{\relax Ghosh}}\ \emph {et~al.}(2015)\citenamefont
  {{\relax Ghosh}}, \citenamefont {{\relax Li}}, \citenamefont {{\relax
  Marchegiani}},\ and\ \citenamefont {{\relax
  Marchesoni}}}]{biophys:Ghosh.2015}%
  \BibitemOpen
  \bibfield  {author} {\bibinfo {author} {\bibfnamefont {P.~K.}\ \bibnamefont
  {{\relax Ghosh}}}, \bibinfo {author} {\bibfnamefont {Y.}~\bibnamefont
  {{\relax Li}}}, \bibinfo {author} {\bibfnamefont {G.}~\bibnamefont {{\relax
  Marchegiani}}}, \ and\ \bibinfo {author} {\bibfnamefont {F.}~\bibnamefont
  {{\relax Marchesoni}}},\ }\href@noop {} {\bibfield  {journal} {\bibinfo
  {journal} {J. Chem. Phys.}\ }\textbf {\bibinfo {volume} {143}},\ \bibinfo
  {pages} {211101} (\bibinfo {year} {2015})}\BibitemShut {NoStop}%
\bibitem [{\citenamefont {{\relax Ni}}\ \emph {et~al.}(2013)\citenamefont
  {{\relax Ni}}, \citenamefont {{\relax Stuart}},\ and\ \citenamefont {{\relax
  Dijkstra}}}]{biophys:Ni.2013}%
  \BibitemOpen
  \bibfield  {author} {\bibinfo {author} {\bibfnamefont {R.}~\bibnamefont
  {{\relax Ni}}}, \bibinfo {author} {\bibfnamefont {M.~A.~C.}\ \bibnamefont
  {{\relax Stuart}}}, \ and\ \bibinfo {author} {\bibfnamefont {M.}~\bibnamefont
  {{\relax Dijkstra}}},\ }\href@noop {} {\bibfield  {journal} {\bibinfo
  {journal} {Nature Commun.}\ }\textbf {\bibinfo {volume} {4}},\ \bibinfo
  {pages} {2704} (\bibinfo {year} {2013})}\BibitemShut {NoStop}%
\bibitem [{\citenamefont {{\relax Fily}}\ \emph {et~al.}(2013)\citenamefont
  {{\relax Fily}}, \citenamefont {{\relax Henkes}},\ and\ \citenamefont
  {{\relax Marchetti}}}]{biophys:Fily.2013}%
  \BibitemOpen
  \bibfield  {author} {\bibinfo {author} {\bibfnamefont {Y.}~\bibnamefont
  {{\relax Fily}}}, \bibinfo {author} {\bibfnamefont {S.}~\bibnamefont {{\relax
  Henkes}}}, \ and\ \bibinfo {author} {\bibfnamefont {M.~C.}\ \bibnamefont
  {{\relax Marchetti}}},\ }\href@noop {} {\bibfield  {journal} {\bibinfo
  {journal} {Soft Matter}\ }\textbf {\bibinfo {volume} {10}},\ \bibinfo {pages}
  {2132} (\bibinfo {year} {2013})}\BibitemShut {NoStop}%
\bibitem [{\citenamefont {{\relax Kuan}}\ \emph {et~al.}(2015)\citenamefont
  {{\relax Kuan}}, \citenamefont {{\relax Blackwell}}, \citenamefont {{\relax
  Hough}}, \citenamefont {{\relax Glaser}},\ and\ \citenamefont {{\relax
  Betterton}}}]{biophys:Kuan.2015}%
  \BibitemOpen
  \bibfield  {author} {\bibinfo {author} {\bibfnamefont {H.-S.}\ \bibnamefont
  {{\relax Kuan}}}, \bibinfo {author} {\bibfnamefont {R.}~\bibnamefont {{\relax
  Blackwell}}}, \bibinfo {author} {\bibfnamefont {L.~E.}\ \bibnamefont {{\relax
  Hough}}}, \bibinfo {author} {\bibfnamefont {M.~A.}\ \bibnamefont {{\relax
  Glaser}}}, \ and\ \bibinfo {author} {\bibfnamefont {M.~D.}\ \bibnamefont
  {{\relax Betterton}}},\ }\href@noop {} {\bibfield  {journal} {\bibinfo
  {journal} {Phys. Rev. E}\ }\textbf {\bibinfo {volume} {92}},\ \bibinfo
  {pages} {060501(R)} (\bibinfo {year} {2015})}\BibitemShut {NoStop}%
\bibitem [{\citenamefont {{\relax Berthier}}\ and\ \citenamefont {{\relax
  Kurchan}}(2013)}]{biophys:Berthier.2013}%
  \BibitemOpen
  \bibfield  {author} {\bibinfo {author} {\bibfnamefont {L.}~\bibnamefont
  {{\relax Berthier}}}\ and\ \bibinfo {author} {\bibfnamefont {J.}~\bibnamefont
  {{\relax Kurchan}}},\ }\href@noop {} {\bibfield  {journal} {\bibinfo
  {journal} {Nature Phys.}\ }\textbf {\bibinfo {volume} {9}},\ \bibinfo {pages}
  {310} (\bibinfo {year} {2013})}\BibitemShut {NoStop}%
\bibitem [{\citenamefont {{\relax Berthier}}(2014)}]{biophys:Berthier.2014}%
  \BibitemOpen
  \bibfield  {author} {\bibinfo {author} {\bibfnamefont {L.}~\bibnamefont
  {{\relax Berthier}}},\ }\href@noop {} {\bibfield  {journal} {\bibinfo
  {journal} {Phys. Rev. Lett.}\ }\textbf {\bibinfo {volume} {112}},\ \bibinfo
  {pages} {220602} (\bibinfo {year} {2014})}\BibitemShut {NoStop}%
\bibitem [{\citenamefont {{\relax Levis}}\ and\ \citenamefont {{\relax
  Berthier}}(2015)}]{biophys:Levis.2015}%
  \BibitemOpen
  \bibfield  {author} {\bibinfo {author} {\bibfnamefont {D.}~\bibnamefont
  {{\relax Levis}}}\ and\ \bibinfo {author} {\bibfnamefont {L.}~\bibnamefont
  {{\relax Berthier}}},\ }\href@noop {} {\bibfield  {journal} {\bibinfo
  {journal} {EPL}\ }\textbf {\bibinfo {volume} {111}},\ \bibinfo {pages}
  {60006} (\bibinfo {year} {2015})}\BibitemShut {NoStop}%
\bibitem [{\citenamefont {{\relax Szamel}}\ \emph {et~al.}(2015)\citenamefont
  {{\relax Szamel}}, \citenamefont {{\relax Flenner}},\ and\ \citenamefont
  {{\relax Berthier}}}]{biophys:Szamel.2015}%
  \BibitemOpen
  \bibfield  {author} {\bibinfo {author} {\bibfnamefont {G.}~\bibnamefont
  {{\relax Szamel}}}, \bibinfo {author} {\bibfnamefont {E.}~\bibnamefont
  {{\relax Flenner}}}, \ and\ \bibinfo {author} {\bibfnamefont
  {L.}~\bibnamefont {{\relax Berthier}}},\ }\href@noop {} {\bibfield  {journal}
  {\bibinfo  {journal} {Phys. Rev. E}\ }\textbf {\bibinfo {volume} {91}},\
  \bibinfo {pages} {062304} (\bibinfo {year} {2015})}\BibitemShut {NoStop}%
\bibitem [{\citenamefont {{\relax Mandal}}\ \emph {et~al.}(2016)\citenamefont
  {{\relax Mandal}}, \citenamefont {{\relax Bhuyan}}, \citenamefont {{\relax
  Rao}},\ and\ \citenamefont {{\relax Dasgupta}}}]{biophys:Mandal.2016}%
  \BibitemOpen
  \bibfield  {author} {\bibinfo {author} {\bibfnamefont {R.}~\bibnamefont
  {{\relax Mandal}}}, \bibinfo {author} {\bibfnamefont {P.~J.}\ \bibnamefont
  {{\relax Bhuyan}}}, \bibinfo {author} {\bibfnamefont {M.}~\bibnamefont
  {{\relax Rao}}}, \ and\ \bibinfo {author} {\bibfnamefont {C.}~\bibnamefont
  {{\relax Dasgupta}}},\ }\href@noop {} {\bibfield  {journal} {\bibinfo
  {journal} {Soft Matter}\ }\textbf {\bibinfo {volume} {12}},\ \bibinfo {pages}
  {6268} (\bibinfo {year} {2016})}\BibitemShut {NoStop}%
\bibitem [{\citenamefont {{\relax Takatori}}\ \emph {et~al.}(2014)\citenamefont
  {{\relax Takatori}}, \citenamefont {{\relax Yan}},\ and\ \citenamefont
  {{\relax Brady}}}]{biophys:Takatori.2014}%
  \BibitemOpen
  \bibfield  {author} {\bibinfo {author} {\bibfnamefont {S.~C.}\ \bibnamefont
  {{\relax Takatori}}}, \bibinfo {author} {\bibfnamefont {W.}~\bibnamefont
  {{\relax Yan}}}, \ and\ \bibinfo {author} {\bibfnamefont {J.~F.}\
  \bibnamefont {{\relax Brady}}},\ }\href@noop {} {\bibfield  {journal}
  {\bibinfo  {journal} {Phys. Rev. Lett.}\ }\textbf {\bibinfo {volume}
  {113}},\ \bibinfo {pages} {028103} (\bibinfo {year} {2014})}\BibitemShut
  {NoStop}%
\bibitem [{\citenamefont {{\relax Takatori}}\ and\ \citenamefont {{\relax
  Brady}}(2016)}]{biophys:Takatori.2016}%
  \BibitemOpen
  \bibfield  {author} {\bibinfo {author} {\bibfnamefont {S.~C.}\ \bibnamefont
  {{\relax Takatori}}}\ and\ \bibinfo {author} {\bibfnamefont {J.~F.}\
  \bibnamefont {{\relax Brady}}},\ }\href@noop {} {\bibfield  {journal}
  {\bibinfo  {journal} {Curr. Opin. Colloid Interf. Sci.}\ }\textbf
  {\bibinfo {volume} {21}},\ \bibinfo {pages} {24} (\bibinfo {year}
  {2016})}\BibitemShut {NoStop}%
\bibitem [{\citenamefont {{\relax Yan}}\ and\ \citenamefont {{\relax
  Brady}}(2015{\natexlab{a}})}]{biophys:Yan.2015}%
  \BibitemOpen
  \bibfield  {author} {\bibinfo {author} {\bibfnamefont {W.}~\bibnamefont
  {{\relax Yan}}}\ and\ \bibinfo {author} {\bibfnamefont {J.~F.}\ \bibnamefont
  {{\relax Brady}}},\ }\href@noop {} {\bibfield  {journal} {\bibinfo  {journal}
  {Soft Matter}\ }\textbf {\bibinfo {volume} {11}},\ \bibinfo {pages} {6235}
  (\bibinfo {year} {2015}{\natexlab{a}})}\BibitemShut {NoStop}%
\bibitem [{\citenamefont {{\relax Yan}}\ and\ \citenamefont {{\relax
  Brady}}(2015{\natexlab{b}})}]{biophys:Yan.2015bpre}%
  \BibitemOpen
  \bibfield  {author} {\bibinfo {author} {\bibfnamefont {W.}~\bibnamefont
  {{\relax Yan}}}\ and\ \bibinfo {author} {\bibfnamefont {J.~F.}\ \bibnamefont
  {{\relax Brady}}},\ }\href@noop {} {\enquote {\bibinfo {title} {The force on
  a boundary in active matter},}\ } (\bibinfo {year} {2015}{\natexlab{b}}),\
  \Eprint {http://arxiv.org/abs/cond-mat.soft/1510.07731}
  {cond-mat.soft/1510.07731} \BibitemShut {NoStop}%
\bibitem [{\citenamefont {Kurzthaler}\ \emph {et~al.}(2016)\citenamefont
  {Kurzthaler}, \citenamefont {Leitmann},\ and\ \citenamefont
  {Franosch}}]{biophys:Kurzthaler.2016}%
  \BibitemOpen
  \bibfield  {author} {\bibinfo {author} {\bibfnamefont {C.}~\bibnamefont
  {Kurzthaler}}, \bibinfo {author} {\bibfnamefont {S.}~\bibnamefont
  {Leitmann}}, \ and\ \bibinfo {author} {\bibfnamefont {T.}~\bibnamefont
  {Franosch}},\ }\href@noop {} {\bibfield  {journal} {\bibinfo  {journal} {Sci.
  Rep.}\ }\textbf {\bibinfo {volume} {6}},\ \bibinfo {pages} {36702} (\bibinfo
  {year} {2016})}\BibitemShut {NoStop}%
\bibitem [{\citenamefont {{\relax Wang}}\ and\ \citenamefont {{\relax
  Wolynes}}(2011)}]{biophys:Wang.2011}%
  \BibitemOpen
  \bibfield  {author} {\bibinfo {author} {\bibfnamefont {S.}~\bibnamefont
  {{\relax Wang}}}\ and\ \bibinfo {author} {\bibfnamefont {P.~G.}\ \bibnamefont
  {{\relax Wolynes}}},\ }\href@noop {} {\bibfield  {journal} {\bibinfo
  {journal} {J. Chem. Phys.}\ }\textbf {\bibinfo {volume} {135}},\ \bibinfo
  {pages} {051101} (\bibinfo {year} {2011})}\BibitemShut {NoStop}%
\bibitem [{\citenamefont {{\relax Farage}}\ and\ \citenamefont {{\relax
  Brader}}(2014)}]{biophys:Farage.2014pre}%
  \BibitemOpen
  \bibfield  {author} {\bibinfo {author} {\bibfnamefont {T.~F.~F.}\
  \bibnamefont {{\relax Farage}}}\ and\ \bibinfo {author} {\bibfnamefont
  {J.~M.}\ \bibnamefont {{\relax Brader}}},\ }\href@noop {} {\enquote {\bibinfo
  {title} {Dynamics and rheology of active glasses},}\ } (\bibinfo {year}
  {2014}),\ \Eprint {http://arxiv.org/abs/cond-mat.soft/1403.0928}
  {cond-mat.soft/1403.0928} \BibitemShut {NoStop}%
\bibitem [{\citenamefont {{\relax Farage}}\ \emph {et~al.}(2015)\citenamefont
  {{\relax Farage}}, \citenamefont {{\relax Krinninger}},\ and\ \citenamefont
  {{\relax Brader}}}]{biophys:Farage.2015}%
  \BibitemOpen
  \bibfield  {author} {\bibinfo {author} {\bibfnamefont {T.~F.~F.}\
  \bibnamefont {{\relax Farage}}}, \bibinfo {author} {\bibfnamefont
  {P.}~\bibnamefont {{\relax Krinninger}}}, \ and\ \bibinfo {author}
  {\bibfnamefont {J.~M.}\ \bibnamefont {{\relax Brader}}},\ }\href@noop {}
  {\bibfield  {journal} {\bibinfo  {journal} {Phys. Rev. E}\ }\textbf
  {\bibinfo {volume} {91}},\ \bibinfo {pages} {042310} (\bibinfo {year}
  {2015})}\BibitemShut {NoStop}%
\bibitem [{\citenamefont {{\relax Szamel}}(2015)}]{biophys:Szamel.2016}%
  \BibitemOpen
  \bibfield  {author} {\bibinfo {author} {\bibfnamefont {G.}~\bibnamefont
  {{\relax Szamel}}},\ }\href@noop {} {\bibfield  {journal} {\bibinfo
  {journal} {Phys. Rev. E}\ }\textbf {\bibinfo {volume} {93}},\ \bibinfo
  {pages} {012603} (\bibinfo {year} {2015})}\BibitemShut {NoStop}%
\bibitem [{\citenamefont {{\relax
  Nandi}}(2016{\natexlab{a}})}]{biophys:Nandi.2016pre}%
  \BibitemOpen
  \bibfield  {author} {\bibinfo {author} {\bibfnamefont {S.~K.}\ \bibnamefont
  {{\relax Nandi}}},\ }\href@noop {} {\enquote {\bibinfo {title} {Activity is
  strength: More active systems are stronger glass formers},}\ } (\bibinfo
  {year} {2016}{\natexlab{a}}),\ \Eprint
  {http://arxiv.org/abs/cond-mat.soft/1605.06073} {cond-mat.soft/1605.06073}
  \BibitemShut {NoStop}%
\bibitem [{\citenamefont {{\relax
  Nandi}}(2016{\natexlab{b}})}]{biophys:Nandi.2016bpre}%
  \BibitemOpen
  \bibfield  {author} {\bibinfo {author} {\bibfnamefont {S.~K.}\ \bibnamefont
  {{\relax Nandi}}},\ }\href@noop {} {\enquote {\bibinfo {title} {Steady state
  of active systems is characterized by unique effective temperature},}\ }
  (\bibinfo {year} {2016}{\natexlab{b}}),\ \Eprint
  {http://arxiv.org/abs/cond-mat.soft/1607.04478} {cond-mat.soft/1607.04478}
  \BibitemShut {NoStop}%
\bibitem [{\citenamefont {{\relax Götze}}(2009)}]{glass-theory:Goetze.2009}%
  \BibitemOpen
  \bibfield  {author} {\bibinfo {author} {\bibfnamefont {W.}~\bibnamefont
  {{\relax Götze}}},\ }\href@noop {} {\emph {\bibinfo {title} {Complex
  Dynamics of Glass-Forming Liquids}}}\ (\bibinfo  {publisher} {Oxford
  University Press},\ \bibinfo {year} {2009})\BibitemShut {NoStop}%
\bibitem [{\citenamefont {{\relax Bayer}}\ \emph {et~al.}(2007)\citenamefont
  {{\relax Bayer}}, \citenamefont {{\relax Brader}}, \citenamefont {{\relax
  Ebert}}, \citenamefont {{\relax Fuchs}}, \citenamefont {{\relax Lange}},
  \citenamefont {{\relax Maret}}, \citenamefont {{\relax Schilling}},
  \citenamefont {{\relax Sperl}},\ and\ \citenamefont {{\relax
  Wittmer}}}]{glass-theory:Bayer.2007}%
  \BibitemOpen
  \bibfield  {author} {\bibinfo {author} {\bibfnamefont {M.}~\bibnamefont
  {{\relax Bayer}}}, \bibinfo {author} {\bibfnamefont {J.~M.}\ \bibnamefont
  {{\relax Brader}}}, \bibinfo {author} {\bibfnamefont {F.}~\bibnamefont
  {{\relax Ebert}}}, \bibinfo {author} {\bibfnamefont {M.}~\bibnamefont
  {{\relax Fuchs}}}, \bibinfo {author} {\bibfnamefont {E.}~\bibnamefont
  {{\relax Lange}}}, \bibinfo {author} {\bibfnamefont {G.}~\bibnamefont
  {{\relax Maret}}}, \bibinfo {author} {\bibfnamefont {R.}~\bibnamefont
  {{\relax Schilling}}}, \bibinfo {author} {\bibfnamefont {M.}~\bibnamefont
  {{\relax Sperl}}}, \ and\ \bibinfo {author} {\bibfnamefont {J.~P.}\
  \bibnamefont {{\relax Wittmer}}},\ }\href@noop {} {\bibfield  {journal}
  {\bibinfo  {journal} {Phys. Rev. E}\ }\textbf {\bibinfo {volume} {76}},\
  \bibinfo {pages} {011508} (\bibinfo {year} {2007})}\BibitemShut {NoStop}%
\bibitem [{\citenamefont {{\relax Ding}}\ \emph {et~al.}(2015)\citenamefont
  {{\relax Ding}}, \citenamefont {{\relax Feng}}, \citenamefont {{\relax
  Jiang}},\ and\ \citenamefont {{\relax Hou}}}]{biophys:Ding.2015pre}%
  \BibitemOpen
  \bibfield  {author} {\bibinfo {author} {\bibfnamefont {H.}~\bibnamefont
  {{\relax Ding}}}, \bibinfo {author} {\bibfnamefont {M.}~\bibnamefont {{\relax
  Feng}}}, \bibinfo {author} {\bibfnamefont {H.}~\bibnamefont {{\relax
  Jiang}}}, \ and\ \bibinfo {author} {\bibfnamefont {Z.}~\bibnamefont {{\relax
  Hou}}},\ }\href@noop {} {\enquote {\bibinfo {title} {Nonequilibrium glass
  transition in mixtures of active-passive particles},}\ } (\bibinfo {year}
  {2015}),\ \Eprint {http://arxiv.org/abs/cond-mat.soft/1506.02754}
  {cond-mat.soft/1506.02754} \BibitemShut {NoStop}%
\bibitem [{\citenamefont {{\relax Szamel}}(2014)}]{biophys:Szamel.2014}%
  \BibitemOpen
  \bibfield  {author} {\bibinfo {author} {\bibfnamefont {G.}~\bibnamefont
  {{\relax Szamel}}},\ }\href@noop {} {\bibfield  {journal} {\bibinfo
  {journal} {Phys. Rev. E}\ }\textbf {\bibinfo {volume} {90}},\ \bibinfo
  {pages} {012111} (\bibinfo {year} {2014})}\BibitemShut {NoStop}%
\bibitem [{\citenamefont {{\relax Marconi}}\ and\ \citenamefont {{\relax
  Maggi}}(2015)}]{biophys:Marconi.2015}%
  \BibitemOpen
  \bibfield  {author} {\bibinfo {author} {\bibfnamefont {U.~M.~B.}\
  \bibnamefont {{\relax Marconi}}}\ and\ \bibinfo {author} {\bibfnamefont
  {C.}~\bibnamefont {{\relax Maggi}}},\ }\href@noop {} {\bibfield  {journal}
  {\bibinfo  {journal} {Soft Matter}\ }\textbf {\bibinfo {volume} {11}},\
  \bibinfo {pages} {8768} (\bibinfo {year} {2015})}\BibitemShut {NoStop}%
\bibitem [{\citenamefont {{\relax Sadjadi}}\ \emph {et~al.}(2015)\citenamefont
  {{\relax Sadjadi}}, \citenamefont {{\relax Shaebani}}, \citenamefont {{\relax
  Rieger}},\ and\ \citenamefont {{\relax Santen}}}]{biophys:Sadjadi.2015}%
  \BibitemOpen
  \bibfield  {author} {\bibinfo {author} {\bibfnamefont {Z.}~\bibnamefont
  {{\relax Sadjadi}}}, \bibinfo {author} {\bibfnamefont {M.~R.}\ \bibnamefont
  {{\relax Shaebani}}}, \bibinfo {author} {\bibfnamefont {H.}~\bibnamefont
  {{\relax Rieger}}}, \ and\ \bibinfo {author} {\bibfnamefont {L.}~\bibnamefont
  {{\relax Santen}}},\ }\href@noop {} {\bibfield  {journal} {\bibinfo
  {journal} {Phys. Rev. E}\ }\textbf {\bibinfo {volume} {91}},\ \bibinfo
  {pages} {062715} (\bibinfo {year} {2015})}\BibitemShut {NoStop}%
\bibitem [{\citenamefont {{\relax Levis}}\ and\ \citenamefont {{\relax
  Berthier}}(2014)}]{biophys:Levis.2014}%
  \BibitemOpen
  \bibfield  {author} {\bibinfo {author} {\bibfnamefont {D.}~\bibnamefont
  {{\relax Levis}}}\ and\ \bibinfo {author} {\bibfnamefont {L.}~\bibnamefont
  {{\relax Berthier}}},\ }\href@noop {} {\bibfield  {journal} {\bibinfo
  {journal} {Phys. Rev. E}\ }\textbf {\bibinfo {volume} {89}},\ \bibinfo
  {pages} {062301} (\bibinfo {year} {2014})}\BibitemShut {NoStop}%
\bibitem [{\citenamefont {{\relax Flenner}}\ \emph {et~al.}(2016)\citenamefont
  {{\relax Flenner}}, \citenamefont {{\relax Szamel}},\ and\ \citenamefont
  {{\relax Berthier}}}]{biophys:Flenner.2016pre}%
  \BibitemOpen
  \bibfield  {author} {\bibinfo {author} {\bibfnamefont {E.}~\bibnamefont
  {{\relax Flenner}}}, \bibinfo {author} {\bibfnamefont {G.}~\bibnamefont
  {{\relax Szamel}}}, \ and\ \bibinfo {author} {\bibfnamefont {L.}~\bibnamefont
  {{\relax Berthier}}},\ }\href@noop {} {\enquote {\bibinfo {title} {The
  nonequilibrium glassy dynamics of self-propelled particles},}\ } (\bibinfo
  {year} {2016}),\ \Eprint {http://arxiv.org/abs/cond-mat.soft/1606.00641}
  {cond-mat.soft/1606.00641} \BibitemShut {NoStop}%
\bibitem [{\citenamefont {{\relax Reichhardt}}\ and\ \citenamefont {{\relax
  Olson Reichhardt}}(2014)}]{biophys:Reichhardt.2014}%
  \BibitemOpen
  \bibfield  {author} {\bibinfo {author} {\bibfnamefont {C.}~\bibnamefont
  {{\relax Reichhardt}}}\ and\ \bibinfo {author} {\bibfnamefont {C.~J.}\
  \bibnamefont {{\relax Olson Reichhardt}}},\ }\href@noop {} {\bibfield
  {journal} {\bibinfo  {journal} {Soft Matter}\ }\textbf {\bibinfo {volume}
  {10}},\ \bibinfo {pages} {7502} (\bibinfo {year} {2014})}\BibitemShut
  {NoStop}%
\bibitem [{\citenamefont {{\relax Franosch}}\ \emph {et~al.}(1997)\citenamefont
  {{\relax Franosch}}, \citenamefont {{\relax Fuchs}}, \citenamefont {{\relax
  Götze}}, \citenamefont {{\relax Mayr}},\ and\ \citenamefont {{\relax
  Singh}}}]{glass-theory:Franosch.1997c}%
  \BibitemOpen
  \bibfield  {author} {\bibinfo {author} {\bibfnamefont {T.}~\bibnamefont
  {{\relax Franosch}}}, \bibinfo {author} {\bibfnamefont {M.}~\bibnamefont
  {{\relax Fuchs}}}, \bibinfo {author} {\bibfnamefont {W.}~\bibnamefont
  {{\relax Götze}}}, \bibinfo {author} {\bibfnamefont {M.~R.}\ \bibnamefont
  {{\relax Mayr}}}, \ and\ \bibinfo {author} {\bibfnamefont {A.~P.}\
  \bibnamefont {{\relax Singh}}},\ }\href@noop {} {\bibfield  {journal}
  {\bibinfo  {journal} {Phys. Rev. E}\ }\textbf {\bibinfo {volume} {56}},\
  \bibinfo {pages} {5659} (\bibinfo {year} {1997})}\BibitemShut {NoStop}%
\bibitem [{\citenamefont {{\relax Kämmerer}}\ \emph
  {et~al.}(1997)\citenamefont {{\relax Kämmerer}}, \citenamefont {{\relax
  Kob}},\ and\ \citenamefont {{\relax
  Schilling}}}]{glass-theory:Kaemmerer.1997}%
  \BibitemOpen
  \bibfield  {author} {\bibinfo {author} {\bibfnamefont {S.}~\bibnamefont
  {{\relax Kämmerer}}}, \bibinfo {author} {\bibfnamefont {W.}~\bibnamefont
  {{\relax Kob}}}, \ and\ \bibinfo {author} {\bibfnamefont {R.}~\bibnamefont
  {{\relax Schilling}}},\ }\href@noop {} {\bibfield  {journal} {\bibinfo
  {journal} {Phys. Rev. E}\ }\textbf {\bibinfo {volume} {56}},\ \bibinfo
  {pages} {5450} (\bibinfo {year} {1997})}\BibitemShut {NoStop}%
\bibitem [{\citenamefont {{\relax
  Schilling}}(2002)}]{glass-theory:Schilling.2002}%
  \BibitemOpen
  \bibfield  {author} {\bibinfo {author} {\bibfnamefont {R.}~\bibnamefont
  {{\relax Schilling}}},\ }\href@noop {} {\bibfield  {journal} {\bibinfo
  {journal} {Phys. Rev. E}\ }\textbf {\bibinfo {volume} {65}},\ \bibinfo
  {pages} {051206} (\bibinfo {year} {2002})}\BibitemShut {NoStop}%
\bibitem [{\citenamefont {{\relax Tailleur}}\ and\ \citenamefont {{\relax
  Cates}}(2008)}]{biophys:Tailleur.2008}%
  \BibitemOpen
  \bibfield  {author} {\bibinfo {author} {\bibfnamefont {J.}~\bibnamefont
  {{\relax Tailleur}}}\ and\ \bibinfo {author} {\bibfnamefont {M.~E.}\
  \bibnamefont {{\relax Cates}}},\ }\href@noop {} {\bibfield  {journal}
  {\bibinfo  {journal} {Phys. Rev. Lett.}\ }\textbf {\bibinfo {volume}
  {100}},\ \bibinfo {pages} {218103} (\bibinfo {year} {2008})}\BibitemShut
  {NoStop}%
\bibitem [{\citenamefont {{\relax Fuchs}}\ and\ \citenamefont {{\relax
  Cates}}(2002)}]{rheology:Fuchs.2002c}%
  \BibitemOpen
  \bibfield  {author} {\bibinfo {author} {\bibfnamefont {M.}~\bibnamefont
  {{\relax Fuchs}}}\ and\ \bibinfo {author} {\bibfnamefont {M.~E.}\
  \bibnamefont {{\relax Cates}}},\ }\href@noop {} {\bibfield  {journal}
  {\bibinfo  {journal} {Phys. Rev. Lett.}\ }\textbf {\bibinfo {volume}
  {89}},\ \bibinfo {pages} {248304} (\bibinfo {year} {2002})}\BibitemShut
  {NoStop}%
\bibitem [{\citenamefont {{\relax Gazuz}}\ \emph {et~al.}(2009)\citenamefont
  {{\relax Gazuz}}, \citenamefont {{\relax Puertas}}, \citenamefont {{\relax
  Voigtmann}},\ and\ \citenamefont {{\relax Fuchs}}}]{rheology:Gazuz.2009}%
  \BibitemOpen
  \bibfield  {author} {\bibinfo {author} {\bibfnamefont {I.}~\bibnamefont
  {{\relax Gazuz}}}, \bibinfo {author} {\bibfnamefont {A.~M.}\ \bibnamefont
  {{\relax Puertas}}}, \bibinfo {author} {\bibfnamefont {T.}~\bibnamefont
  {{\relax Voigtmann}}}, \ and\ \bibinfo {author} {\bibfnamefont
  {M.}~\bibnamefont {{\relax Fuchs}}},\ }\href@noop {} {\bibfield  {journal}
  {\bibinfo  {journal} {Phys. Rev. Lett.}\ }\textbf {\bibinfo {volume}
  {102}},\ \bibinfo {pages} {248302} (\bibinfo {year} {2009})}\BibitemShut
  {NoStop}%
\bibitem [{\citenamefont {{\relax Enculescu}}\ and\ \citenamefont {{\relax
  Stark}}(2011)}]{biophys:Enculescu.2011}%
  \BibitemOpen
  \bibfield  {author} {\bibinfo {author} {\bibfnamefont {M.}~\bibnamefont
  {{\relax Enculescu}}}\ and\ \bibinfo {author} {\bibfnamefont
  {H.}~\bibnamefont {{\relax Stark}}},\ }\href@noop {} {\bibfield  {journal}
  {\bibinfo  {journal} {Phys. Rev. Lett.}\ }\textbf {\bibinfo {volume}
  {107}},\ \bibinfo {pages} {058301} (\bibinfo {year} {2011})}\BibitemShut
  {NoStop}%
\bibitem [{\citenamefont {{\relax Tailleur}}\ and\ \citenamefont {{\relax
  Cates}}(2009)}]{biophys:Tailleur.2009}%
  \BibitemOpen
  \bibfield  {author} {\bibinfo {author} {\bibfnamefont {J.}~\bibnamefont
  {{\relax Tailleur}}}\ and\ \bibinfo {author} {\bibfnamefont {M.~E.}\
  \bibnamefont {{\relax Cates}}},\ }\href@noop {} {\bibfield  {journal}
  {\bibinfo  {journal} {EPL}\ }\textbf {\bibinfo {volume} {86}},\ \bibinfo
  {pages} {60002} (\bibinfo {year} {2009})}\BibitemShut {NoStop}%
\bibitem [{\citenamefont {{\relax Bialké}}\ \emph {et~al.}(2012)\citenamefont
  {{\relax Bialké}}, \citenamefont {{\relax Speck}},\ and\ \citenamefont
  {{\relax Löwen}}}]{biophys:Bialke.2012}%
  \BibitemOpen
  \bibfield  {author} {\bibinfo {author} {\bibfnamefont {J.}~\bibnamefont
  {{\relax Bialké}}}, \bibinfo {author} {\bibfnamefont {T.}~\bibnamefont
  {{\relax Speck}}}, \ and\ \bibinfo {author} {\bibfnamefont {H.}~\bibnamefont
  {{\relax Löwen}}},\ }\href@noop {} {\bibfield  {journal} {\bibinfo
  {journal} {Phys. Rev. Lett.}\ }\textbf {\bibinfo {volume} {108}},\ \bibinfo
  {pages} {168301} (\bibinfo {year} {2012})}\BibitemShut {NoStop}%
\bibitem [{\citenamefont {{J\"{u}lich Supercomputing Centre}}(2016)}]{jureca}%
  \BibitemOpen
  \bibfield  {author} {\bibinfo {author} {\bibnamefont {{J\"{u}lich
  Supercomputing Centre}}},\ }\href {\doibase 10.17815/jlsrf-2-121} {\bibfield
  {journal} {\bibinfo  {journal} {Journal of large-scale research facilities}\
  }\textbf {\bibinfo {volume} {2}} (\bibinfo {year} {2016}),\
  10.17815/jlsrf-2-121}\BibitemShut {NoStop}%
\end{thebibliography}%
\bibliographystyle{apsrev4-1}

\end{document}